\documentclass[acmsmall]{acmart}

\usepackage{amsmath}

\usepackage{amssymb}
\usepackage{amsfonts}
\usepackage{algorithmic}
\usepackage{textcomp}
\usepackage{xcolor}
\usepackage{float}
\usepackage{algorithm}
\usepackage{booktabs}
\usepackage{subfigure}
\usepackage{pifont}
\usepackage{graphicx}
\usepackage{epstopdf}
\usepackage{epsfig}
\usepackage{multirow}
\usepackage{makecell}
\usepackage{bbding}
\usepackage{threeparttable}
\usepackage{tikz}
\usepackage{wasysym}
\AtBeginDocument{%
  \providecommand\BibTeX{{%
    \normalfont B\kern-0.5em{\scshape i\kern-0.25em b}\kern-0.8em\TeX}}}

\setcopyright{acmcopyright}
\copyrightyear{2018}
\acmYear{2018}
\acmDOI{XXXXXXX.XXXXXXX}

\acmJournal{JACM}
\acmVolume{37}
\acmNumber{4}
\acmArticle{111}
\acmMonth{8}

\begin{document}

\title{Backdoor Attacks against Voice Recognition Systems: A Survey}

\author{Baochen Yan}
\affiliation{
    \institution{The State Key Lab of ISN, Xidian University}
    \country{China}
}
\email{bcyan@stu.xidian.edu.cn}

\author{Jiahe Lan}
\affiliation{
  \institution{The State Key Lab of ISN, Xidian University}
  \country{China}
}
\email{jhlan16@stu.xidian.edu.cn}

\author{Zheng Yan}
\authornote{Corresponding Author}
\affiliation{
    \institution{The State Key Lab of ISN, Xidian University}
    \country{China}
}
\email{zyan@xidian.edu.cn}

\thanks{This work is supported in part by the National Natural Science Foundation of China under Grant 62072351; supported by the Fundamental Research Funds for the Central Universities, No. YJSJ23007; in part by the Key Research Project of Shaanxi Natural Science Foundation under Grant 2023-JC-ZD-35; in part by the open research project of ZheJiang Lab under grant 2021PD0AB01; in part by the 111 Project under Grant B16037, and in part by the Fundamental Research Funds for the Central Universities under grant YJSJ23007.}


\renewcommand{\shortauthors}{B.C. Yan, et al.}

\begin{abstract}
    Voice Recognition Systems (VRSs) employ deep learning for speech recognition and speaker recognition. They have been widely deployed in various real-world applications, from intelligent voice assistance to telephony surveillance and biometric authentication. However, prior research has revealed the vulnerability of VRSs to backdoor attacks, which pose a significant threat to the security and privacy of VRSs. Unfortunately, existing literature lacks a thorough review on this topic. This paper fills this research gap by conducting a comprehensive survey on backdoor attacks against VRSs. We first present an overview of VRSs and backdoor attacks, elucidating their basic knowledge. Then we propose a set of evaluation criteria to assess the performance of backdoor attack methods. Next, we present a comprehensive taxonomy of backdoor attacks against VRSs from different perspectives and analyze the characteristic of different categories. After that, we comprehensively review existing attack methods and analyze their pros and cons based on the proposed criteria. Furthermore, we review classic backdoor defense methods and generic audio defense techniques. Then we discuss the feasibility of deploying them on VRSs. Finally, we figure out several open issues and further suggest future research directions to motivate the research of VRSs security.

\end{abstract}



\received{20 February 2007}
\received[revised]{12 March 2009}
\received[accepted]{5 June 2009}

\maketitle

\section{Introduction}


As a crucial human-computer interface, Voice Recognition Systems (VRSs) have been widely deployed in classical and emerging intelligent devices, such as smartphones, laptops, and smart home devices, to assist human-device interaction. Speech recognition systems ~\cite{wang2022improving,zhang2022wenetspeech} and speaker recognition systems ~\cite{zhang2022volere,zhang2022livoauth,yan2022voicesketch} are two common VRSs. The former utilizes semantic features of human speech to convert speech into text. The latter aid in identifying the speaker via personal voice features of human speech. 
Such interfaces not only facilitate our daily lives ~\cite{ammari2019music} but also improve accessibility for groups such as the elderly and the visually impaired ~\cite{pradhan2018accessibility, pradhan2020use}, make devices without screens accessible ~\cite{porcheron2018voice} and make user authentication seamless ~\cite{rui2018survey}.

The unprecedented performance exhibited by VRSs ~\cite{xie2019utterance} is inseparable from the rapid development of deep learning. Nevertheless, training a powerful deep learning model requires a substantial amount of training data and extensive computing resources, making it unaffordable for ordinary users and even small businesses ~\cite{zhuang2020comprehensive}. Utilizing third-party resources, such as open-source datasets ~\cite{Voxceleb}, open-source pre-trained models ~\cite{qiu2020pre}, and third-party model training platforms ~\cite{colab} is a great way to alleviate the problem of insufficient resources and help ordinary users or small businesses enjoy the convenience brought by deep learning. However, this provides new attack surfaces for attackers. Untrusted third parties may provide malicious datasets and pre-trained models, and even maliciously manipulate the entire model training process, posing significant security risks. One typical threat is backdoor attacks ~\cite{badnet}.

\begin{figure*}[htb]
    \centering
    \includegraphics[width=135mm]{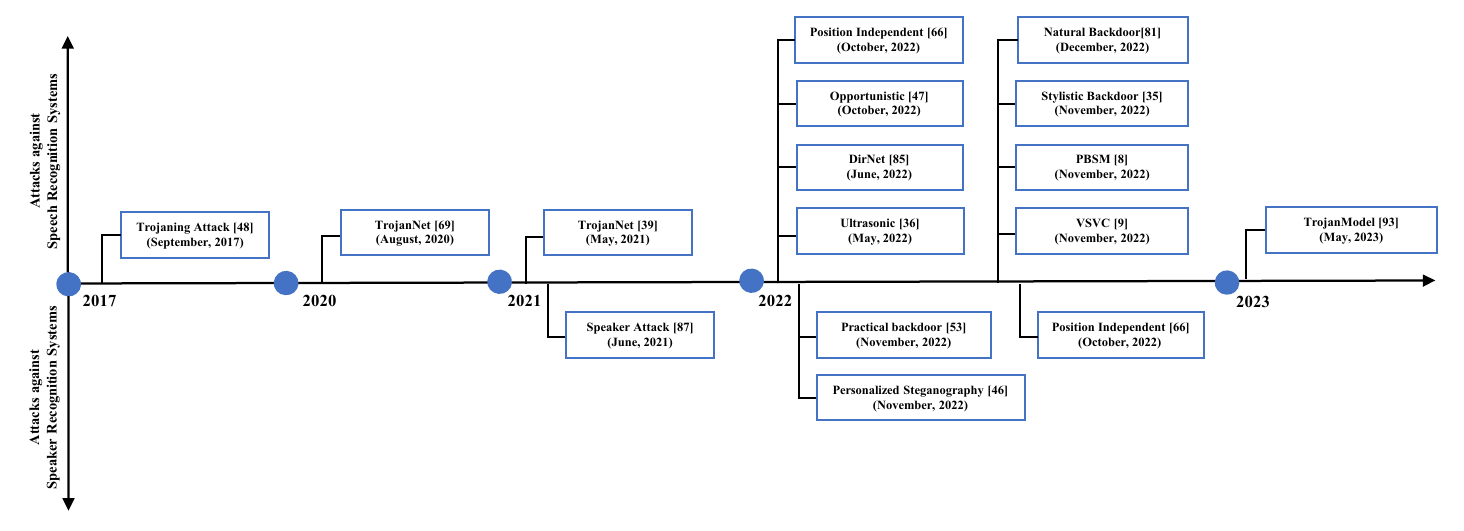}
    \caption{A Timeline of Backdoor Attacks against Voice Recognition Systems.}
    \label{time}
\end{figure*}

Backdoor attacks aim to embed one or more hidden backdoors into Deep Neural Networks (DNNs) ~\cite{dnn}. The infected model performs well on benign samples, while the hidden backdoor in the infected model can be activated by a predefined trigger, resulting in unexpected predictions. Backdoor attacks were initially proposed and studied in the image domain ~\cite{badnet}, and gradually migrated into the audio domain ~\cite{an_novel, trojaning_ndss}, especially in the recent three years.
The development history of backdoor attacks in VRSs is presented in Fig \ref{time}.
The upper part shows typical backdoor attacks against speech recognition systems and the bottom part shows typical backdoor attacks against speaker recognition systems before July 2023.

Several researchers have surveyed backdoor attacks and defenses in different fields ~\cite{review1,review2,review5,review6,review7,review4,10082866}. 
Researchers in ~\cite{review1}, ~\cite{review7}, ~\cite{review5} and ~\cite{review6} conducted comprehensive surveys on backdoor attacks and defenses in the vision domain or both vision and text domains. However, they ignored audio backdoor attacks and defenses. 
To the best of our knowledge, Gao et al. \cite{review2} first investigated audio backdoor attacks in their survey although they mainly focused on the vision and text domain. Notably, they assessed the cross-domain effectiveness of backdoor defenses using an evaluation metric - generalization. Specifically, most existing backdoor defenses are designed for the vision domain, and generalization can be used to evaluate the effectiveness of these defenses in the audio and text domains.
Zong et al. ~\cite{review4} conducted the first survey that only focused on audio backdoor attacks and defenses. However, this survey is incomplete, as it only covers two backdoor attacks and two defenses. Furthermore, it only focuses on backdoor attacks against speech recognition systems and ignores backdoor attacks against speaker recognition systems.
Recently, Ge et al. ~\cite{10082866} surveyed both poisoning and backdoor attacks against VRSs. Nevertheless, they ignored the associated countermeasures. 
In order to clarify the difference and novelty of our survey, a detailed comparison of our survey and highly related surveys is given in Table \ref{CompareSurveys}. 
We can observe that our survey stands out as the only one that proposes a comprehensive taxonomy of audio backdoor attacks and defenses. Additionally, we conduct a thorough review of audio backdoor attacks and analyze the feasibility of employing existing backdoor defenses against VRSs.
The security of VRSs has been a hot topic of research, and other excellent surveys have investigated other attacks on the audio domain, such as evasion attacks ~\cite{lan2022adversarial}, poisoning attacks ~\cite{miller2020adversarial} and spoofing attacks ~\cite{wu2015spoofing}. Therefore, a comprehensive survey on the recent advances in audio backdoor attacks and defenses is urgently expected in the literature to help researchers and practitioners to understand the current state of the research, open issues, and future research directions toward promoting the research on secure and trustworthy VRSs.

\begin{table*}[htbp]
  \centering
  \caption{Compassion of Our Survey with Other Existing Surveys}
  \label{CompareSurveys}
    \renewcommand\arraystretch{1.1}
    \resizebox{\linewidth}{!}
    {
    \begin{tabular}{c|c|c|c|c|c|c|c}
    \hline
    \hline
    \multicolumn{1}{c|}{ \textbf{Paper}} & \ \, \textbf{Year} \, \, & \textbf{Field} & \ding{172}  & \ding{173}     & \ding{174}     & \ding{175}    & \ding{176}  \\
    \hline
    ~\cite{review1} & 2020  & Vision & \quad \CIRCLE \quad\quad  &\quad \quad\quad \quad&\quad \CIRCLE \quad\quad &\quad\quad\quad\quad & \quad\quad\quad\quad \\
    \hline
    ~\cite{review5}  & 2020  & Vision, Text  &   &  & \CIRCLE &   &   \\
    \hline
    ~\cite{review2} & 2020  & Vision, Text, Audio  & \CIRCLE  & \Circle & \CIRCLE & \Circle & \CIRCLE \\
    \hline
    ~\cite{review6}  & 2022  & Vision, Text & \CIRCLE &  \Circle & \CIRCLE &  & \\
    \hline
    ~\cite{review7}  & 2022  &  Vision 
     &    & \CIRCLE & \CIRCLE  &  &  \\
    \hline
    ~\cite{review4}  & 2022  &  Audio
     &  &   &  & \Circle & \Circle \\
    \hline
     ~\cite{10082866}  & 2023  &  Audio
     & \Circle & \CIRCLE &  & \CIRCLE &  \\
    \hline
    This paper & 2023  & Audio & \CIRCLE  &  \CIRCLE  &   \CIRCLE & \CIRCLE & \CIRCLE \\
    \hline
    \multicolumn{8}{c}{\CIRCLE: Fully supported; \Circle: Partially supported; None: Not supported;}\\
    \multicolumn{8}{c}{\ding{172}: Compare backdoor attacks with other attacks} \\
    \multicolumn{8}{c}{\ding{173}: Propose a set of criteria that a sound backdoor attack should meet} \\
    \multicolumn{8}{c}{\ding{174}: Present a taxonomy of backdoor attacks and defenses, \ding{175}: Review audio backdoor attacks} \\
    \multicolumn{8}{c}{\ding{176}: Discuss the feasibility of applying existing backdoor defenses} \\
    \multicolumn{8}{c}{and generic audio defense to VRSs } \\
    \hline
    \hline
    \end{tabular}%
    }
\end{table*}%

In this paper, we intend to conduct a comprehensive survey on audio backdoor attacks and defenses. 
We first introduce the basic knowledge of VRSs and backdoor attacks. At the same time, we analyze and discuss the similarities and differences between backdoor attacks and other adversarial attacks, including evasion attacks, poisoning attacks, and inference attacks.
Next, we propose a set of criteria that a sound audio backdoor attack should is expected to meet.
After that, we propose taxonomies of audio backdoor attacks from different perspectives, such as target system and attack property, and discuss the characteristics of each category.
Based on the taxonomies and the proposed evaluation criteria, we comprehensively review audio backdoor attacks and analyze their pros and cons.
In addition, considering the lack of audio backdoor defenses, we discuss whether and how existing advanced image backdoor defenses can be transferred to the audio domain, as well as some generic audio defense techniques.
In the end, we shed light on several open issues and suggest future research directions.
To summarize, the main contributions of this paper are as below: 
\begin{itemize}
    \item We propose a set of evaluation criteria that should be satisfied by an audio backdoor attack.
    \item We conduct an in-depth review on existing audio backdoor attacks against VRSs by employing the proposed evaluation criteria to analyze their pros and cons.
    \item We discuss potential audio backdoor defenses, deriving from image backdoor defenses and generic audio defense techniques.
    \item We point out a list of open issues and further propose future research directions to promote the development of trustworthy VRSs.
\end{itemize}

The remainder of this survey is organized as follows.
In the next section, we make an overview of VRSs and backdoor attacks and compare different adversarial attacks. In Section III, we propose a set of criteria for evaluating the performance of audio backdoor attacks. Section IV presents taxonomies of audio backdoor attacks, followed by a thorough review on existing works with deep-insight analysis by employing our proposed criteria. In Section V, we discuss potential audio backdoor defenses. On the basis of the literature review, we highlight open issues and suggest future research directions in Section VI. Finally, we draw a conclusion in the last section.

\section{Background Knowledge}
This section first provides an introduction to the basics of the two VRSs, namely, speech recognition systems and speaker recognition systems. Then, we briefly introduced several common audio features used in VRSs. Subsequently, we present the concept of backdoor attacks. Following that, we outline the similarities and differences between backdoor attacks and other attacks in deep learning.

\subsection{Voice Recognition Systems}

\begin{figure*}[t]
    \centering 
    \subfigure[An Overview of Speech Recognition Systems.]{ 
    \begin{minipage}{14cm}
    \centering
    \includegraphics[width=70mm]{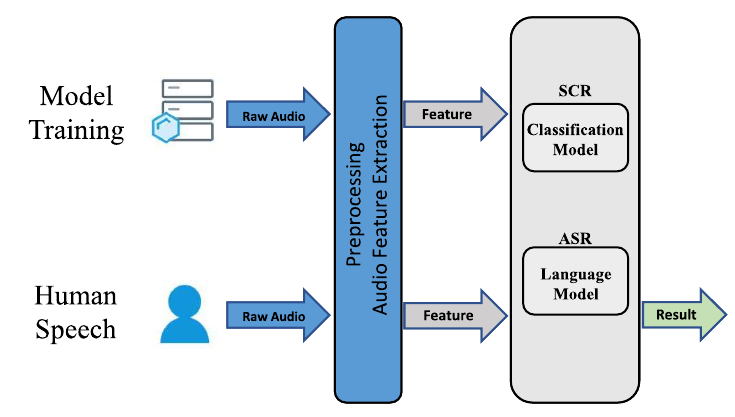} 
    \end{minipage}
    }
    \subfigure[An Overview of Speaker Recognition systems.]{
    \begin{minipage}{14cm}
    \centering 
    \includegraphics[width=105mm]{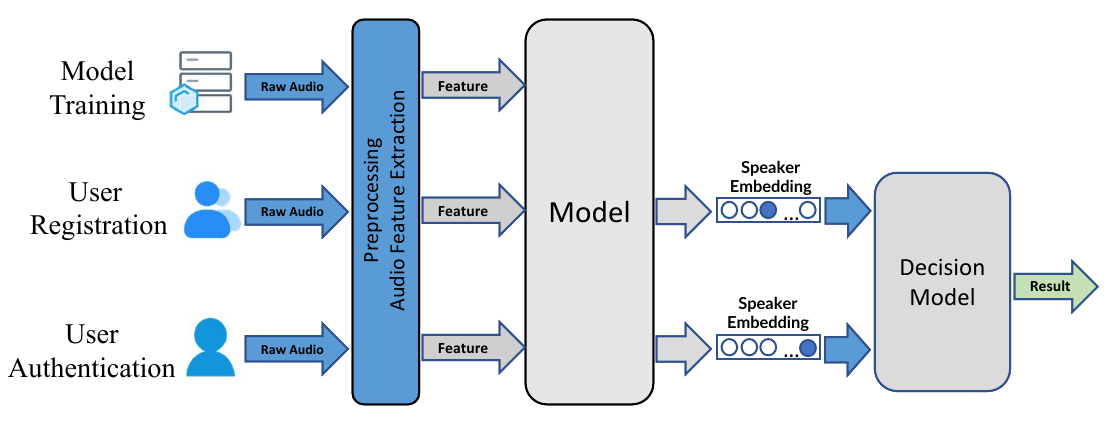}
    \end{minipage}
    }
    \caption{General Frameworks of DNN-based Voice Recognition Systems.} 
    \label{frameworks} 
\end{figure*}

A VRS is a technological system that involves the processing, analysis, and recognition of speech signals. The main objective of this system is to process human audio input using computer algorithms and models to extract useful information for further analysis and application.
Speech recognition and speaker recognition systems are the two typical types of VRSs. They are widely used in our daily life, such as intelligent voice assistants ~\cite{assistants, hoy2018alexa}, speech transcription ~\cite{speech_transcription, ahmed2020preech}, voiceprint authentication~\cite{zhang2022volere, zhang2022livoauth} and access control~\cite{access_control}.

\subsubsection{Speech Recognition Systems}
Speech recognition systems ~\cite{specch11} are designed to identify the content of speech. Fig. \ref{frameworks} (a) depicts the workflow of a speech recognition system. Firstly, a preprocessing component is employed to enhance the quality of an audio signal, commonly utilizing functions such as frequency filtering and noise reduction. Secondly, an acoustic feature extractor captures the acoustic feature vector, such as the spectrum and spectrogram, of the preprocessed audio. Lastly, a prediction model utilizes the acoustic feature vector to generate the final textual representation.


Depending on different purposes, speech recognition systems can be divided into Speech Command Recognition (SCR) and Automatic Speech Recognition (ASR).
SCR is designed to accurately recognize specific audio commands. Since audio commands are typically simple and have predefined word lists or limited vocabulary, SCR often relies on simple classification models. SCR is commonly employed in automotive voice controls and interactions with voice assistants.
Differently, ASR is designed to transcribe and comprehend natural human language speech, surpassing the capabilities of SCR. ASR employs sophisticated language models and techniques to effectively handle the intricacies of natural language, encompassing grammar, syntax, and context. ASR is commonly used in various applications such as transcription services and voice search.

\subsubsection{Speaker Recognition Systems}
Speaker recognition systems ~\cite{speaker11, yan2016usable} attempt to identify speakers through their speech ~\cite{rui2018survey}. Fig. \ref{frameworks} (b) shows the workflow of a speaker recognition system, including three stages: training stage, enrollment stage, and recognition stage.
In the training stage, a corpus is used to train a speaker feature extractor $E(\cdot)$, which extracts a feature vector (also known as an embedding) representing the identity of the speaker from the speaker's speech.
After that, $n$ speakers, i.e., a group of speakers $G = {speaker_1, speaker_2,..., speaker_n}$, enroll in the speaker recognition systems, and their speeches are represented as $s_1, s_2,...,s_n$. The speaker recognition system calculates and stores feature vectors of enrolled speakers, i.e., $E(s_1), E(s_2),..., E(s_n)$. 
Lastly, in the recognition stage, the speaker recognition system identifies an unknown input $x$ via a decision module, and the decision module outputs $D(x)$.
There are three common sub-tasks of speaker recognition systems: Speaker Verification (SV), Close-Set Identification (CSI), and Open-Set Identification (OSI).

For an OSI system, in the recognition stage, it first calculates the similarity between the embedding of $x$, i.e., $E(x)$, and the embeddings of enrolled speakers one by one. The similarity between $E(x)$ and $E(s_i)$ is represented as $Sim(E(x), E(s_i))$, where $i = 1,2,...,n$ and $Sim(\cdot)$ is a similarity calculation function, such as cosine similarity. And then, the system determines whether $x$ is uttered by one of the enrolled speakers or none of them, according to similarities and a predefined similarity threshold $\theta$. Intuitively, the system classifies the input voice $x$ as the $speaker_i$ if and only if the similarity $Sim(E(x), E(s_i))$ is the largest one among all the enrolled speakers and not less than the threshold $\theta$. The formal expression of OSI is shown in Equation 1.

\begin{equation}
    D(x)=\left\{\begin{array}{l}
    \underset{i \in G}{\operatorname{argmax}} \ Sim(E(x),E(s_i)), \quad \text { if } \underset{i \in G}{\operatorname{max}} \ Sim(E(x),E(s_i)) > \theta \\
    \text {reject}, \quad \text {otherwise}
    \end{array}\right.
\end{equation}

A CSI system accomplishes a similar task as an OSI system. However, it never rejects any inputs, i.e., an input voice $x$ must be classified as one of the enrolled speakers. The formal expression of CSI is shown in Equation 2.
\begin{equation}
    D(x)=\underset{i \in G}{\operatorname{argmax}} \ Sim(E(x),E(s_i))
\end{equation} 

As for an SV system, the unknown input audio $x$ claims its owner is $speaker_i$. Thus, the system determines whether $x$ is uttered by $speaker_i$ according to the similarity between the embedding of $x$ and $speaker_i$ and the predefined threshold $\theta$. The formal expression of SV is shown in Equation 3.

\begin{equation}
    D(x)=\left\{\begin{array}{ll} 
    i, \quad \text { if } Sim(E(x),E(s_i)) > \theta \\ 
    \text {reject}, \quad \text {otherwise}
    \end{array}\right.
\end{equation}

\subsection{Audio Features}
Audio features offer diverse representations and descriptions of various aspects of an audio signal and play a crucial role in both speech recognition and speaker recognition. This subsection presents commonly used audio features extracted from the time domain, frequency domain, and time-frequency domain.

\subsubsection{Time Domain Features}
Time domain features are derived from the analysis of the audio signal in the time domain. Waveform is the most commonly used time-domain feature.

\begin{itemize}
    \item \textbf{Waveform}: A waveform ~\cite{waveform} is considered the most straightforward audio feature, as it represents the amplitude variation of an audio signal over time.
\end{itemize}

\subsubsection{Frequency Domain Features}
Frequency domain features are obtained by performing Fourier transform or other spectral transform methods on the speech signal. Spectrum is a commonly used frequency-domain feature.


\begin{itemize}
    \item \textbf{Spectrum}: A spectrum ~\cite{Spectrum} refers to the distribution of energy or power across different frequencies in an audio signal. It provides information about the frequency content and characteristics of the audio signal.
\end{itemize}

\subsubsection{Time-frequency Domain Features}
Time-frequency domain features combine both time and frequency information to describe the variations of the audio signal over time and frequency. Common time-frequency domain features include spectrogram and Mel-Frequency Cepstral Coefficient (MFCC).
\begin{itemize}
    \item \textbf{Spectrogram}: A spectrogram ~\cite{Spectrogram} is a two-dimensional representation of the spectrum over time. It provides a detailed view of the frequency content of an audio signal and how it changes over time. 
    
 
    \item \textbf{MFCC}: A MFCC ~\cite{mfcc} is a kind of advanced audio feature that considers the human perception of frequencies. It offers a compact and robust representation that is less sensitive to noise and channel variations, making it widely favored in speech-processing applications.
\end{itemize}

\subsection{Backdoor Attacks}

\begin{figure*}[htb]
    \centering
    \includegraphics[width=130mm]{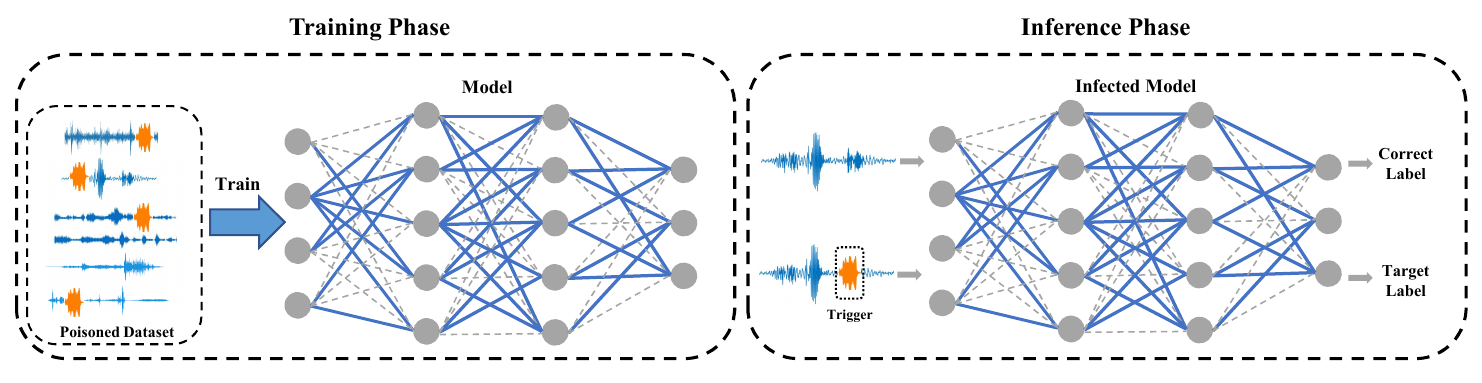}
    \caption{An Illustration of Poisoned-Based Backdoor Attacks.}
    \label{Illustration}
\end{figure*}

A backdoor attack aims to insert one or more hidden backdoors into a DNN model so that the attacked model (also known as the \emph{infected model}) performs well on \emph{benign samples} while producing attacker-desired predictions for samples with triggers (i.e., \emph{poisoned samples}).


Figure \ref{Illustration} demonstrates a typical poisoned-based backdoor attack.
Initially, an attacker constructs a poisoned training dataset by poisoning a training dataset. 
Specifically, he randomly selects some training samples and adds triggers to them. After that, an attacker modifies the labels of these training samples to the target labels.
Subsequently, when a model is trained on the poisoned training dataset, it is regarded to be injected with a hidden backdoor.
During the inference phase, the infected model is capable of accurately recognizing benign samples.
Conversely, poisoned samples, i.e., samples with triggers, are classified as the target class.

Poisoning training datasets is the most common way to deploy backdoor attacks.
In addition, some works directly modify the parameters or the loss function of models to insert backdoors~\cite{weight_modify, bagdasaryan2021blind}.
Notably, these attack methods can be executed without requiring access to the training dataset, facilitating their generalization.
There is a growing number of studies paying attention to non-poisoned-based attack methods, which are more flexible and difficult to detect~\cite{li2021deeppayload}.

\subsection{Comparison between Backdoor Attacks and Other Adversarial Attacks on DNNs}

\begin{table}[htbp]\footnotesize
    \centering
    \caption{Comparison between Backdoor Attack and Other Adversarial Attacks}
    \label{Adversarial}
    \renewcommand\arraystretch{1}
    \resizebox{\linewidth}{!}
    {
    \begin{tabular}{m{2cm}<{\centering}|m{3cm}<{\centering}|m{2cm}<{\centering}|m{2cm}<{\centering}|m{3cm}<{\centering}}
        \hline
        \hline
        \textbf{Attack Category} & \multicolumn{1}{c|}{\textbf{Attack Goal}} & \textbf{Broken Security Property} & \textbf{Attack Phase} & \multicolumn{1}{c}{\textbf{Vulnerability}} \\
        \hline
        Evasion Attack & Misclassify adversarial samples & Integrity & Inference phase & Sensitivity of input \\
        \hline
        Non-Targeted Poisoning Attack & Influence the overall performance of a model & Integrity & Training phase & Dependence on training data \\
        \hline
        Targeted Poisoning Attack & Misclassify samples of a specific class & Integrity & Training phase & Dependence on training data \\
        \hline
        Inference Attack & Leak the parameters or training data of a target model & Confidentiality & Inference phase & Risk of privacy leakage \\
        \hline
        Backdoor Attack & Misclassify poisoned samples & Integrity & Training and Inference phase & Plasticity of DNNs \\
        \hline
        \hline
    \end{tabular}%
    }
    \label{tab:addlabel}%
\end{table}%

Adversarial attacks refer to the attacks regarding DNN-based machine learning systems ~\cite{A1}.
Different adversarial attacks target different security properties related to the integrity or confidentiality of DNNs.
The former consists of evasion attacks, poisoning attacks, and backdoor attacks. The latter includes inference attacks.
Overall, the backdoor attack represents a distinctive type of adversarial attack targeting DNNs, presenting unique challenges and advantages compared to other attack types. Therefore, comprehending the differences between backdoor attacks and other forms of attacks is crucial for understanding backdoor attacks and other adversarial attacks in DNNs.
This subsection briefly compares backdoor attacks and other attacks against DNNs in terms of attack goal, broken security property, attack phase, and attack method, as shown in Table \ref{Adversarial}.
\subsubsection{Evasion Attacks}


The evasion attack ~\cite{lan2022adversarial} aims to manipulate DNNs during the inference phase by introducing adversarial perturbations into its input samples. 
This manipulation leads to the model producing incorrect outputs with a high level of confidence.
It is also known as the adversarial example attack.
The evasion attack compromises the integrity of DNNs and is launched in the inference phase.

\subsubsection{Poisoning Attacks}
The poisoning attack ~\cite{poisoningattack} refers to the deliberate act of injecting malicious samples into the training dataset of DNNs, resulting in compromised model performance.	
Depending on the difference in impact, the poisoning attack can be divided into non-targeted poisoning attack and targeted poisoning attack.
The goal of the non-targeted poisoning attack ~\cite{untargetedpoisoning} is to make the model unusable, significantly impacting its overall performance. In other words, the attack goal is to cause the DNNs to consistently produce incorrect predictions.
In contrast, the targeted poisoning attack ~\cite{targetedpoisoning} causes the infected model to misclassify a specific class (i.e., a source label) as another target class (i.e., a target label), without affecting the overall performance of the infected model. In other words, it only misclassifies a particular target class.	
The poisoning attack is implemented during the training phase and compromises the integrity of DNNs.


\subsubsection{Inference Attacks}
The inference attack ~\cite{inferenceattack} is a data mining technique used to illicitly acquire information about a target by analyzing data.
The membership inference attack ~\cite{memeberinference} aims to infer the existence of specific data samples in a target model's training dataset, while the model inference attack ~\cite{modelinference} aims to deduce the attribute values of target data in the training dataset based on model outputs.
Inference attacks primarily target the confidentiality of DNNs and are launched during the model's inference phase.	

\subsubsection{A Comparison between Backdoor Attacks and Other Adversarial Attacks}
Backdoor attacks share a common attack goal with evasion attacks and targeted poisoning attacks, which is to intentionally misclassify specific input samples.
Unlike evasion attacks, which necessitate the creation of unique adversarial perturbations for each input, backdoor attacks attempt to utilize a specific trigger on arbitrary samples to manipulate the malicious behavior of DNNs.
In contrast to targeted poisoning attacks, where the misclassification behavior is beyond the attacker's control in the inference phase, backdoor attacks can be manipulated by using a trigger and offer greater flexibility.	

In terms of the broken security property, backdoor attacks align with poisoning attacks and evasion attacks as they all compromise the integrity of DNNs. Conversely, inference attacks target the confidentiality of DNNs.
And in terms of the attack phase, evasion attacks, poisoning attacks, and inference attacks take place either during the training phase or the inference phase of DNNs. Only backdoor attacks have the capability to be performed in both phases.	

Different adversarial attacks target different vulnerabilities of DNNs.
Evasion attacks leverage the inherent sensitivity of DNNs to input, where even minor perturbations can result in significant alterations in the model's output.
Poisoning attacks exploit the dependence of DNNs on training data. In poisoning attacks, an attacker introduces poisoned samples into the training dataset, which can disrupt the model's training process, leading to the acquisition of incorrect associations or the emergence of inconsistent behavior.
Inference attacks exploit the potential for privacy leakage in DNNs. An attacker can scrutinize the model outputs and deduce internal information about the model, including details like training data, model weights, and more, based on outputs. Such attacks capitalize on the implicit information contained in the model's output, enabling the attacker to reverse engineer the model's properties and internal structure through careful observation.	
Lastly, backdoor attacks capitalize on the plasticity of DNNs. An attacker clandestinely can introduce a malicious backdoor either during the training phase or into an already trained model.
When an infected model detects a trigger in input samples, backdoors will be activated and produce malicious behaviors. 
In Table \ref{Adversarial}, we compare different attacks from different aspects as we mentioned above.

\section{Evaluation Criteria}
\begin{figure*}[tb]
    \centering
    \includegraphics[width=135mm]{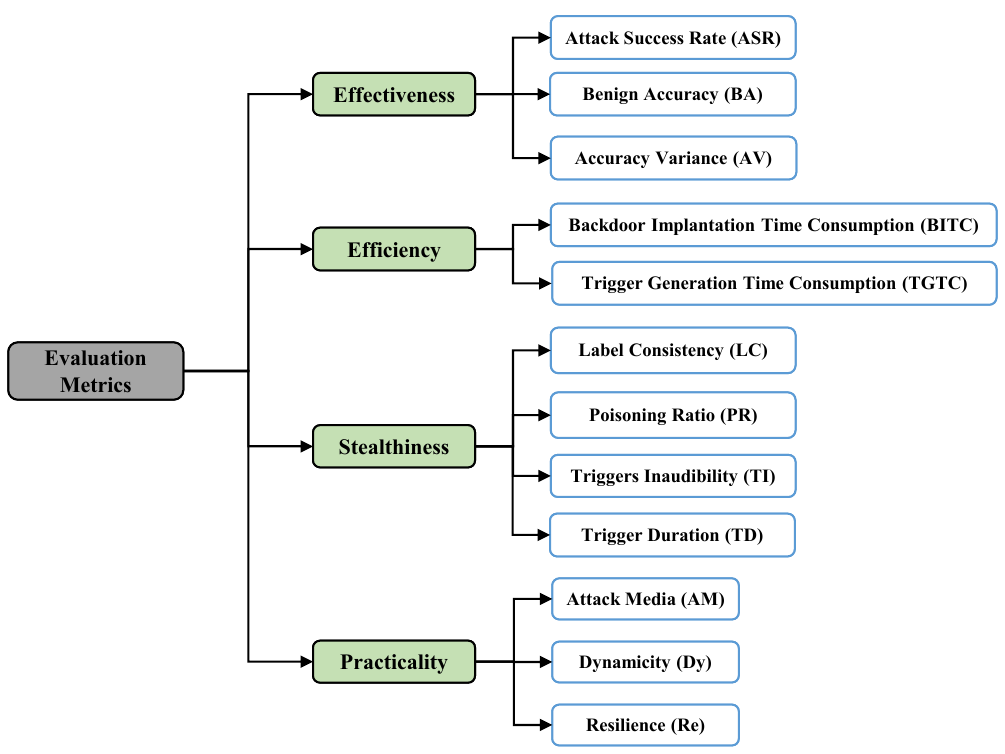}
    \caption{Evaluation Criteria on Backdoor Attacks against VRSs}
    \label{Evaluation}
\end{figure*}

In this section, we propose a set of evaluation criteria on the backdoor attacks against VRSs in terms of four aspects: effectiveness, efficiency, stealthiness, and practicability, as shown in Figure \ref{Evaluation}.

\subsection{Effectiveness}
Effectiveness is used to measure the attack capability of backdoor attacks and is the most important criterion of backdoor attacks.
An effective backdoor attack should achieve a high attack success rate while performing well on benign samples. 
To evaluate the effectiveness of a backdoor attack, we propose the following three quantitative metrics.
\begin{itemize}
\item \textbf{Attack Success Rate (ASR)}: It denotes the proportion of attacked samples that are successfully predicted as the target label by the infected model. 
ASR stands as the most intuitive metric for evaluating the effectiveness of backdoor attacks. A higher ASR indicates superior effectiveness.

 \item \textbf{Benign Accuracy (BA)}: It refers to the accuracy of benign samples predicted by the infected model. BA is an indirect metric for evaluating the impact of backdoor attacks on model performance. 
However, evaluating the effectiveness of a backdoor attack based solely on BA is insufficient unless compared to the accuracy of an original model.	

\item \textbf{Accuracy Variance (AV)}: It represents the absolute difference between the accuracy of a benign model and the BA of the corresponding infected model. AV is an extension of BA and is a direct metric for quantifying the impact of a backdoor attack on a model. Excessive AV is easily detected by humans and leads to attack exposure. Therefore, the lower the AV of a backdoor attack, the better its effectiveness.
\end{itemize}

\subsection{Efficiency}
Efficiency represents the cost associated with launching a backdoor attack.
While ensuring the effectiveness of the backdoor attack, it is crucial to achieve high efficiency.	
To evaluate the efficiency of a backdoor attack, we rely on the following measures.
\begin{itemize}
\item \textbf{Backdoor Implantation Time Consumption (BITC)}: Depending on the method of backdoor implantation into a target model, we categorize BITC into three levels: High, Medium, and Low.
Specifically, when an attacker needs to retrain an infected model from scratch, we consider that BITC is high.
When an attacker utilizes transfer learning or another method to retrain an existing target model, we consider that BITC is medium.
Particularly, if an attacker directly modifies model parameters instead of training with a poisoned dataset, BITC is considered low.

\item \textbf{Trigger Generation Time Consumption (TGTC)}: Similar to our evaluation approach for BITC, TGTC is assessed based on two levels: High and Low.
When trigger audio is directly and readily available, its TGTC is classified as low.
if an attacker requires alternative methods like an optimization-based approach to generate the trigger, TGTC is considered high.
\end{itemize} 

\subsection{Stealthiness}
To ensure that an infected model can be trained without detection, the poisoned sample inserted into a training dataset during the model training phase should be imperceptible to humans.
Similarly, the input samples containing triggers should remain imperceptible to humans during the inference phase.	
To evaluate the stealthiness of a backdoor attack, we propose the following four metrics.
\begin{itemize}
\item \textbf{Label Consistency (LC)}: It means that the content and the label of poisoned samples should be consistent in human perception ~\cite{labelConsistency}. In a traditional backdoor attack, an attacker modifies the label of the poisoned sample to a target label, and the poisoned sample seems to be mislabeled, resulting in a mislabeled sample that is easily noticeable during the manual inspection. Therefore, LC is a basic requirement for a stealthy backdoor attack. 
It is worth noting that LC is only designed to evaluate poisoned-based backdoor attacks.

\item \textbf{Poisoning Ratio (PR)}: It refers to the percentage of poisoned samples in a training dataset. A lower PR reduces the likelihood of detecting the attack. Similar to LC, PR is also only used to evaluate poisoned-based backdoor attacks.

\item \textbf{Trigger Inaudibility (TI)}: TI quantifies the perceptibility of a trigger to humans.
We divide trigger inaudibility into three levels, i.e., high, medium, and low.
A low-level inaudible trigger typically consists of attacker-specified noise and is readily noticeable by humans.	
A medium-level inaudible trigger exhibits some degree of stealthiness, but can still be perceived by highly attentive humans, such as with natural sounds or meticulously calculated triggers.	
A high-level inaudible trigger typically involves ultrasound, which surpasses the frequency range of human hearing (i.e., 20Hz$ \sim $20kHz), rendering it completely imperceptible to humans.	
As TI level of the trigger increases, the stealthiness of the backdoor attack improves.

\item \textbf{Trigger Duration (TD)}: It refers to the duration of the audio being used as a trigger. Generally speaking, the shorter of TD, the less likely it is to be detected by humans, and the better of stealthiness.	
\end{itemize}

\subsection{Practicality}
Practicability refers to the ability of an attack method to be used in the real world.
We rely on the following three metrics to evaluate the practicability of a backdoor attack.

\begin{itemize}
\item \textbf{Attack Medium (AM)}: There are two common attack media including over-line and over-air.
In an over-line attack, the trigger audio is directly transmitted to the model in the form of a waveform or spectrogram audio file, making it the easiest method to execute due to its lossless transmission.
In an over-air attack, the trigger audio needs to be played in the real world.	
An over-air attack is more challenging to implement compared to an over-line attack due to environmental noise and sound attenuation during transmission.
Although the over-air attack is more difficult to launch, it is closer to the real world and its practicability is better.

\item \textbf{Dynamicity (Dy)}: It refers to a trigger that can be added into any position of audio to successfully launch an attack, suitable for dynamic attack scenarios.
In a static attack, triggers are consistently injected at a fixed temporal position within audio used in the training and interface phases.
However, in real-world attack scenarios, it is hard for an attacker to determine when a victim starts to speak.
Different from static attacks, dynamic attacks do not require any form of synchronization between the trigger and the target audio waveform.
Backdoor attacks that satisfy Dy have greater practicality in real-world attack scenarios.

\item \textbf{Resilience (Re)}: Resilience is used to measure the ability of a backdoor attack to bypass defenses ~\cite{fine-pruning, neural_cleanse}. A backdoor attack with Re has better practicality.

\end{itemize}

\section{Backdoor Attacks against VRSs}
In recent years, an increasing number of backdoor attacks against VRSs have emerged. 
In this section, we conducted a comprehensive search for relevant papers in ACM, IEEE, Springer, Elsevier, and Arxiv databases by using keywords: backdoor attack, speech recognition, speaker recognition, etc. from 2018 to 2023.
Based on the query results, we first present a comprehensive taxonomy of backdoor attacks against VRSs and then conduct a comprehensive review of existing backdoor attacks against speech recognition systems and speaker recognition systems according to the evaluation criteria proposed in Section 3. Finally, we summarize our review with a comparison and discussion.

\subsection{Classifications of Backdoor Attacks against VRSs}
\begin{figure*}[tb]
    \centering
    \includegraphics[width=120mm]{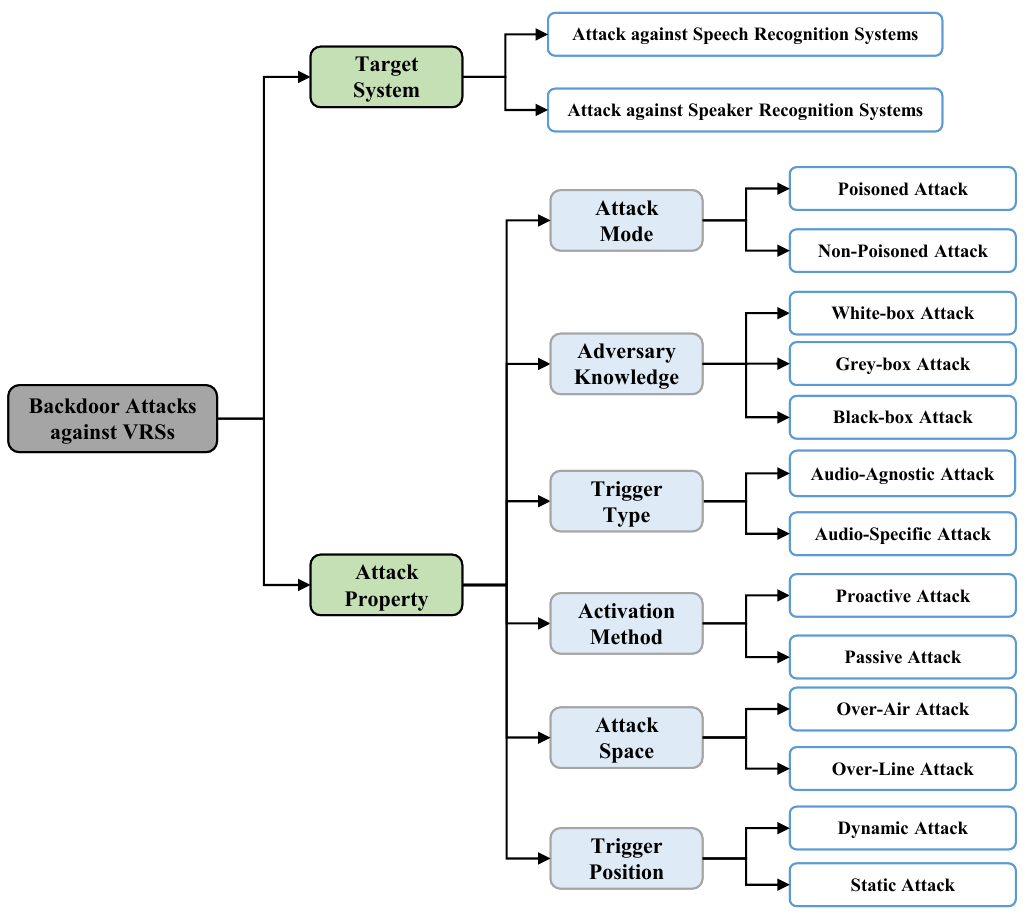}
    \caption{Different Classifications of Backdoor Attacks against VRSs}
    \label{Classifications}
\end{figure*}

Backdoor attacks against VRSs can be classified according to different criteria in terms of attack targets and attack properties. A detailed classification is shown in Figure \ref{Classifications}.

\subsubsection{Classification Based on Target Systems}
Based on the type of attack target systems, we can divide backdoor attacks against VRSs into attacks against speech recognition systems and attacks against speaker recognition systems.
\begin{itemize}
    \item \textbf{Attacks against Speech Recognition Systems}: Backdoor attacks against speech recognition systems aim to induce incorrect recognition of speech content.
    Furthermore, it can be further divided into attacks against SCR and attacks against ASR.
    \item \textbf{Attacks against Speaker Recognition Systems}: Backdoor attacks against speaker recognition systems aim to enable unauthorized users to bypass authentication successfully. Furthermore, it can be further divided into attacks against OSI, attacks against CSI, and attacks against SV.
\end{itemize}

\subsubsection{Classification Based on Attack Properties}
Based on the different attack properties of an attacker, we can classify the backdoor attacks against VRSs into different classes.
    \begin{itemize}
        \item \textbf{Attack Mode:} Based on attack modes, backdoor attacks against VRSs can be divided into poisoning attacks and non-poisoning attacks.
            \begin{itemize}
                \item \textbf{Poisoning Attack}: In a poisoning attack, a backdoor is embedded by poisoning a training dataset.
                This method is widely used and straightforward for executing backdoor attacks. Nevertheless, training or retraining a target model with a manipulated dataset can be computationally demanding and time-consuming, mainly due to the complexity of DNNs.
                
                \item \textbf{Non-Poisoning Attack}: In a non-poisoning attack, an attacker directly alters the model's parameters or loss function to incorporate backdoors into the target model.
                However, directly modifying the model's structure or parameters can potentially cause greater harm to the model's performance compared to the poisoning attack.

            \end{itemize}
    
        \item \textbf{Attacker Capability:} Based on the capabilities of attackers, backdoor attacks against VRSs can be divided into white-box attacks, grey-box attacks, and black-box attacks.
        Table \ref{models} demonstrates the varying capabilities of attackers in these three types of attacks.	
            \begin{table*}[htbp]
              \centering
              \caption{Three Different Threat Models of Backdoor Attacks.}
              \label{models}
                \renewcommand\arraystretch{1}
                \resizebox{0.7\textwidth}{!}
                {
                    \begin{tabular}{c|c|c|c}
                    \hline
                    \hline
                    &\multicolumn{1}{c|}{Learning Task} & \multicolumn{1}{c|}{Target Model} & \multicolumn{1}{c}{Training Dataset} \\
                    \hline
                    White-box & \CIRCLE  & \CIRCLE    & \CIRCLE  \\
                    \hline
                    Grey-box  & \CIRCLE  &\multicolumn{2}{c}{\LEFTcircle} \\
                    \hline
                    Black-box & \CIRCLE  & \Circle    & \Circle  \\
                    \hline
                    \multicolumn{4}{c}{\CIRCLE: Full Knowledge, \LEFTcircle: Partial Knowledge, \Circle: None Knowledge} \\ 
                    \hline
                    \hline
                    \end{tabular}%
                } 
            \end{table*}%
            \begin{itemize}
                \item \textbf{White-box Attack}: The white-box attack occurs when an attacker possesses complete knowledge of the learning task, the target model, and the training dataset, even if the attack only exploits a portion of it. This attack represents the most favorable scenario for the attacker.	
                \item \textbf{Grey-box Attack}: The grey-box attack takes place when an attacker possesses knowledge of the learning task but only has partial information about the target model and the training dataset.	 \item \textbf{Black-box Attack}: The black-box attack occurs when an attacker has only knowledge of the learning task. This attack represents the most challenging scenario for the attacker.
            \end{itemize}
    
        \item \textbf{Trigger Type:} Based on trigger types, the backdoor attacks against VRSs can be divided into audio-agnostic backdoor attacks and audio-specific backdoor attacks, depending on whether the same trigger is used for different input audio samples. 
            \begin{itemize}
                \item \textbf{Audio-agnostic Attack}: In an audio-agnostic backdoor attack, all audio samples share the same trigger. Formally, let $x$ be a clean input and $\epsilon$ be an audio trigger. $\mathcal{F}(\cdot)$ stands for a process function to implant triggers. Audio-agnostic attacks can be described as:
                    \begin{equation}
                    \mathcal{F}(x, \epsilon)=x+\epsilon.
                    \end{equation}
                The audio-agnostic backdoor attack is relatively easy to be implemented. However, a fixed trigger pattern may result in easy attack detection by human beings.
                
                \item \textbf{Audio-specific Attack}: In an audio-specific backdoor attack, a trigger generates function $\mathcal{S}_\epsilon$ is used to generate different trigger $\epsilon$ for different input audio sample $x$ and $\mathcal{F}(\cdot)$ is used to implant different triggers. The audio-specific attack can be described as:
                    \begin{equation}
                    \mathcal{F}\left(x, \mathcal{S}_\epsilon\right)=\mathcal{S}_\epsilon(x).
                    \end{equation}
                The audio-specific backdoor attack tends to be more resilient than the audio-agnostic backdoor attack. However, it has to generate different triggers for different voice audios, so its time consumption is relatively high, thus hard to achieve real-time attacks. In addition, an attacker needs to know the attacked audio in advance, which limits its deployment.
            \end{itemize}

        \item \textbf{Activation Method:} Based on the activation methods of backdoor attacks, the backdoor attacks against VRSs can be divided into proactive and passive attacks. 
            \begin{itemize}
                \item  \textbf{Proactive Attack}: In a proactive attack, an attacker can determine when a backdoor is triggered. Whether a backdoor is activated in an infected model can be controlled by attackers.
                \item \textbf{Passive Attack}: The passive attack is characterized by being passively triggered and opportunistically invoked, thus more harmful than the proactive attack. It may be more easily detected by human beings.
            \end{itemize}
            
        \item \textbf{Attack Space:} Based on attack spaces, the backdoor attacks against VRSs can be divided into over-air attacks and over-line attacks.
            \begin{itemize}
            \item \textbf{Over-line Attack}: In an over-line attack, the trigger is implanted into the target audio by directly modifying the audio feature. This attack is relatively straightforward to execute due to lossless transmission.	                
            \item \textbf{Over-air Attack}: In an over-air attack, a trigger needs to be played to activate the backdoors. This attack is more challenging to implement due to environmental noise and sound attenuation.
            \end{itemize}
            
        \item \textbf{Trigger Position:} Based on trigger positions, the backdoor attacks against VRSs can be divided into dynamic attacks and static attacks.
            \begin{itemize}
                \item \textbf{Dynamic Attack}: In a dynamic attack, a trigger can be inserted anywhere in the target audio to activate a backdoor. This approach is applicable in reality as an attacker cannot control when a victim starts speaking. 
                \item \textbf{Static Attack}: In a static attack, a trigger can only be inserted at a fixed position, such as the beginning or the end of a target audio. In real-world scenarios, static attacks generally don't work.
            \end{itemize}
    \end{itemize}

\subsection{Backdoor Attacks against Speech Recognition Systems}

\subsubsection{Audio-Agnostic Backdoor Attacks}
To the best of our knowledge, Liu et al.~\cite{trojaning_ndss} was the first to propose a backdoor attack named Trojan attack on an SCR system with a grey-box setting.
An attacker starts by selecting a trigger mask, such as an apple logo and a square, which is a subset of the input variables that are used to inject a trigger. 
After that, the attacker selects one or a few neurons in an internal layer of a target model as victim neurons. A trojan generation algorithm is used to generate a trigger to make the selected neurons reach maximum values.
The trojan attack retrains the target model directly on a carefully constructed dataset, its PR is not considered.
Meanwhile, the attacker utilizes a reverse engineering algorithm to construct a training dataset since the original training dataset is not available.
Then, the attacker can retrain a portion of the target model by injecting a backdoor into the target model.
By using a 0.1s trigger, the trojan attack achieves a 96.3\% ASR while maintaining a BA rate of 96.8\% and an AV of 2.3\%.
Its BITC is medium since the attacker only retrains a portion of the target model. TGTC is high since the trigger is obtained through optimization.
TI is low since the trigger is noise in human perception.
Trojan attack does not meet LC and Dy. Experiments in an over-air setting and its Re is not discussed.

Tang et al. ~\cite{embarrassingly} proposed a training-free backdoor attack with a grey-box setting called TrojanNet attack, which is the first training-free backdoor attack specifically targeting SCR systems.
First, the attacker utilizes specific black-and-white squares as triggers and trains a tiny malicious backdoor model named TrojanNet. Then, the TrojanNet is integrated into the target model and is responsible for processing trigger regions, such as the bottom right corner of the spectrum. The predictions of the infected model are jointly determined by the original model and the TrojanNet. So, if the TrojanNet detects the trigger, the infected model's predictions are maliciously changed. 
In this attack, the attacker can insert a small number of neurons (TrojanNet) into the target model, but can neither access the training dataset nor retrain the target model, so this attack belongs to the grey-box attack. 
TrojanNet has low BITC and TGTC, thanks to its training-free mechanism and simple trigger pattern, compared with other backdoor attacks. Its Re was taken into account and it can bypass neural cleanse ~\cite{neural_cleanse} defense.
Although the TrojanNet attack is highlighted as effective in the audio domain, no sufficient experimental results were given. Its ASR, BA, AV, TI, and TD were not mentioned in ~\cite{embarrassingly}. Since the TrojanNet attack is a non-poisoning backdoor attack, there is no need to discuss its PR and LC. In addition, this attack uses spectrum images as speech features and black-and-white squares as triggers, which lack real acoustic meaning. For example, a spectrum is modified directly as image features. Thus, this attack cannot be launched over-air. Nevertheless, we still review this paper as it is the first work that attempts a non-poisoning backdoor attack in the audio domain, hoping that additional research can be performed in the community.

Li et al. ~\cite{an_novel} proposed a novel backdoor attack against ASR systems called Trojan Neural Network (TNN) with a white-box setting. 
An attacker first chooses a short piece of audio as an initial trigger. Then the attacker selects a neuron of the target model, which is strongly connected to the trigger region. After that, the trigger is optimized with gradient descent such that the selected neuron are strongly activated. Once the triggers are in place, the attacker poisons the training dataset and fine-tunes a portion of the model. Therefore, its BITC is medium and its TGTC is high. The TNN attack achieves a 100\% ASR while maintaining an 83.5\% BA with only 0.5\% AV. TI is low since the trigger is noise in human perception. Over-air experiments were missed. Other criteria, including PR and TD, are not mentioned. Besides, Dy and Re are also not discussed in ~\cite{an_novel}.

Audible triggers ~\cite{trojaning_ndss} ~\cite{an_novel} could lead to exposure and failure of backdoor attacks. Building upon this observation, Koffas et al.~\cite{ultrasonic_trigger} proposed an ultrasound-based backdoor attack against SCR systems with a grey-box setting. 
They designed an ultrasound clip as a trigger and inject it into training samples. Subsequently, a backdoor is embedded into a victim model through model training. 
With a PR of 0.21\% and a TD of 0.25s, this ultrasound-based attack achieves an 80\% ASR while maintaining a BA rate of 87.47\% and an AV of 0.11\%. 
Since the attackers designate the trigger, its TGTC is low. Its BITC is high as the victim model should be trained from scratch. Since ultrasound is beyond the range of human hearing, the trigger of the ultrasound-based attack is inaudible in human perception, i.e., TI is high.
The ultrasound-based attack employs triggers in the ultrasonic range, outside the human hearing range, ensuring inaudibility in human perception. Therefore, the attack achieves a high TI level.
Furthermore, the authors conducted over-air experiments to demonstrate that the ultrasound-based backdoor attack is effective in the real world. However, this attack does not support Dy and LC. Meanwhile, Re was not discussed, either.

Note that previous works ~\cite{ultrasonic_trigger} require additional devices to play triggers during attacking, which could easily be noticed by humans.
Xin et al. ~\cite{nature_sound} proposed a natural backdoor attack against SCR systems, using ambient sounds, such as birdsong and whistle, as triggers.
They conducted the attack via data poisoning. Ultimately, they successfully attacked three SCR models with a grey-box setting.
By poisoning 5\% of the training dataset and using a 1s trigger, this attack obtains a 99.94\% ASR in a CNN model and a 97.74\% ASR in an RNN model.
Its BA is 92.94\% and 90.78\%, respectively regarding CNN and RNN, while its AV is 0.58\% and 1.46\%.
Furthermore, LC and the over-air attack are taken into consideration in this work. 
Regarding CNN and RNN, this attack respectively achieves more than 90\% ASR and nearly 70\% ASR with a label-consistent setting.
In the over-air attack, it achieves nearly 70\% ASR in a CNN model and 80\% ASR in an RNN model.
Due to the direct addition of ambient sounds to the samples, the TGTC of this attack is low. Conversely, the BITC is high due to the requirement of training the model from scratch.
Its TI is medium, as ambient sounds are less-noticeable but still perceptible to humans.
Besides, this attack does not meet the requirements for Dy and there is no mention of Re.

Similar to ~\cite{nature_sound}, Liu et al. ~\cite{opportunistic} also noticed that ultrasonic-based triggers ~\cite{ultrasonic_trigger} are easily eliminated by pre-processing (such as low-pass filters) and detected by humans with additional equipment.
To overcome these limitations, they proposed an opportunistic backdoor attack with a white-box setting, which can be activated by environmental sounds. Specifically, they first constructed a pool of ambient sounds, including various common ambient sounds. Then, they evaluated the certainty of an attacked model with each environmental sound, and the highest certainty sound is used as a trigger since a high-certainty-trigger indicates that it is very likely to be encountered during user previous interactions and used for model learning. The authors designed a method named performance-oblivious sample binding to carefully select the poisoned samples in a training dataset, reducing the impact of attacks on model performance, i.e., BA. 
In addition, the same audio could be heard differently in different situations. To this end, they proposed two methods: audio amplitude adjustment and noise mixture to obtain a series of trigger variants to augment triggers.
This attack can achieve an ASR of 92.89\% by only poisoning 0.18\% training data. And it can maintain a BA of 96.96\% and its AV is only 2.18\%.
Its BITC is high due to its training process and its TGTC is also high due to trigger selection and augmentation.
Similar to ~\cite{nature_sound} and~\cite{2022new}, this opportunistic backdoor attack can achieve a medium level of TI. But its TD was not mentioned.
Although LC and Dy are not achieved, it is capable of launching over-air attacks and it is resilient to fine-tuning ~\cite{2020Reflection} backdoor defense.
Since nature sounds are everywhere in daily life, this is a passive attack. An attacker has no control over the exact time of backdoor activation.

Inspired by ~\cite{advpulse}, the authors of ~\cite{2022new} noticed that existing backdoor attacks work in static attack scenarios, without considering dynamic attack scenarios in SCR systems. Therefore, they considered dynamic attack scenarios and proposed a position-independent backdoor attack with a white-box setting. Specifically, in the training phase, an attacker inserts an initial trigger at different positions of every training sample, i.e., with different time stamps, and optimizes the poisoned model and the backdoor trigger.
With a 2\% PR and a 0.14s TD, this attack achieves a 99.12\% ASR while maintaining an 85.12\% BA and a 2.88\% AV in a white-box setting.
Its BITC is high due to the from-scratch training process. Its TGTC is also high since the trigger is obtained by solving an optimization problem.
The TI of the trigger is medium because ambient sounds are not suspicious when perceived by humans.
In addition, the authors spent a lot of effort on over-air attacks, taking environmental sound and room impulse response into account. And eventually, they successfully attacked a speech recognition system in an over-air setting. Experiments demonstrated that this attack is able to bypass a defense method -- fine-pruning ~\cite{fine-pruning}. All in all, this work is one of the state-of-the-art works on backdoor attacks against speech recognition systems. 
In particular, this attack is the only one that considers Dy, advancing the development of this community. However, this attack does not hold LC.

Similar to ~\cite{an_novel}, Zong et al. ~\cite{TrojanModel} trained a separate module called TrojanModel for activating backdoors and obtained better performance in ASR systems.
Attackers use more stealthy and natural triggers e.g., a piece of music playing in the background, which can achieve a medium TI and a low TGTC.
This attack is a also non-poisoning backdoor attack with a white-box setting. Thus, its PR is not considered.
TrojanModel is only required to identify triggers and be inserted into the target model. Thus, the BITC is also low.
They achieve an average 97\% ASR in an over-air setting by using Room Impulse Response (RIR). BA, AV, and TI are not mentioned in the experimental part. Besides, Dy and Re are also not discussed.

\begin{table*}[h]
    \centering
    \begin{minipage}[h]{0.83\linewidth}
    \caption{A Comparison of Existing Audio-agnostic Backdoor Attacks against Speech Recognition Systems}
    \label{table1}
    \renewcommand\arraystretch{2}
    \resizebox{\linewidth}{!}
    {
    \begin{tabular}{c|c|c|c|c|c|c|c|c|c|c|c|c|c|c}
    \hline
    \hline
    \multirow{2}[4]{*}{\textbf{Ref}} & \multirow{2}[4]{*}{\textbf{Task}} & \multirow{2}[4]{*}{\textbf{\makecell[c]{Threat \\ Model}}} & \multicolumn{3}{c|}{\textbf{Effectiveness}} & \multicolumn{2}{c|}{\textbf{Efficiency}} & \multicolumn{4}{c|}{\textbf{Stealthiness}} & \multicolumn{3}{c}{\textbf{Practicability}} \\
\cline{4-15}          &       &       & \textbf{ASR} & \textbf{BA} & \textbf{AV} & \textbf{BITC} & \textbf{TGTC} & \textbf{PR} & \textbf{TI} & \textbf{TD} & \textbf{LC} & \textbf{AM} & \textbf{Dy} & \textbf{Re} \\
    \hline
    ~\cite{trojaning_ndss} & SCR & G  & 96.30\% & 96.8\%. & 2.30\% & M & H  & -    & L & 0.1s  & \ding{55}     & Over-line & \ding{55}     & ? \\
    \hline
    ~\cite{embarrassingly} & SCR & G & ? & ? & ? & L & L  & -    & ? & ?    & \ding{55}     & Over-line & -     & \ding{51} \\
    \hline
    ~\cite{an_novel} & ASR & W & 100\% & 83.50\% & 0.50\% & M & H  & ?    & L & ?    & \ding{55}     & Over-line & \ding{55}     & ? \\
    \hline
    ~\cite{ultrasonic_trigger} & SCR & G  & 80\% & 87.47\% & 0.11\% & H  & L   & 0.21\%   & H & 0.25s & \ding{55}     & Over-air & \ding{55}     & ? \\
    \hline
    ~\cite{nature_sound} & SCR & G  & 99.94\%  & 92.25\% & 0.58\% & H  & L   & 5\%   & M & 1s & \ding{51}     & Over-air & \ding{55}     & ? \\
    \hline
    ~\cite{opportunistic} & SCR & W   & 92.89\% & 96.96\% & 1.98\%  & H  & H  & 0.18\%  & M & -    & \ding{55}     & Over-air & \ding{55}     & \ding{51} \\
    \hline
     ~\cite{2022new} & SCR  & W & 99.12\% & 85.12\% & 2.88\% & H  & H  & 2\%   & M & 0.14s & \ding{55}     & Over-air & \ding{51}     & \ding{51} \\
    \hline
     ~\cite{TrojanModel} & ASR  & W & 97\% & ? & ? & L  & L  & -   & M & ? & \ding{55}  & Over-air & \ding{51}     & \ding{51} \\
    \hline
    \hline
    \end{tabular}
    }
    \begin{threeparttable}
    \begin{tablenotes}
        \footnotesize
        \item[*] SCR: Attacks against SCR Systems, ASR: Attacks against ASR Systems
        \item[*] ASR: Attack Success Rate; BA: Benign Accuracy; AV: Accuracy Variance; BITC: Backdoor Implantation Time Consumption; TGTC: Trigger Generation Time Consumption; LC: Label Consistency; PR: Poisoning Ratio; TI: Trigger Inaudibility; TD: Trigger Duration; AM: Attack Medium; Dy: Dynamicity; Re: Resilience
        \item[*] W: White-box; G: Grey-box; L: Low Level; M: Medium Level; H: High Level
        \item[*] \ding{51}: satisfied; \ding{55}: not satisfied; \textbf{-}: not available; ?: not discussed
    \end{tablenotes}
    
  \label{tab:addlabel}%
  \end{threeparttable}
    \end{minipage}
 
\end{table*} 

\subsubsection{Audio-Specific Backdoor Attacks}

Because audio-agnostic backdoor attacks lack flexibility and are easily detected by existing defense methods ~\cite{fine-pruning, strip}, Ye et al. ~\cite{dirnet} proposed the first audio-specific backdoor attack against an SCR system, called DriNet with a white-box setting. Inspired by the generative adversarial network (GAN), they regarded a trigger generation model as a generative model and a target speech recognition model as a discriminative model and jointly train them. Finally, the trigger generation model and the infected target model are obtained simultaneously.
DriNet achieves an ASR of 86.4\% when poisoning 0.5\% of a training dataset and using a 0.2s trigger in a white-box setting.
Besides, it can achieve an ASR of 86.5\% while maintaining BA of at least 84.4\% and its AV is 0.5\%.
DirNet needs to train a dynamic trigger generation network to craft a variety of audio triggers, thus its TGTC is high.
And DirNet needs to train an infected model from scratch, so its BITC is also high.
Unfortunately, they did not control the volume of the triggers, which led to the poisoned sample being easily perceived. Therefore only a low level of TI is achieved. Besides, LC was not achieved.
As for practicability, the DirNet attack can bypass neural cleanse backdoor defense ~\cite{neural_cleanse}.
However, Dy was not considered and over-air attacks cannot be implemented.

Cai et al. ~\cite{pbsm} found that most of the previous works ~\cite{trojaning_ndss} neglected the stealthiness of the trigger to human ears and existing inaudible triggers ~\cite{ultrasonic_trigger} have significant limitations.
To overcome the above shortcomings, they designed a new backdoor attack with a grey-box setting, named pitch boosting and sound masking (PBSM) ~\cite{pbsm}.
They first boost an audio pitch to provide a stealth space for high-amplitude signal insertion. After that, they search each audio to find the location of the highest amplitude segment of the entire audio. With the two steps above, they generated high-pitch audio and obtained the location of the highest amplitude segment in this audio. Finally, they generated a millisecond short-duration high-amplitude signal and insert it before or after the highest amplitude segment in this audio. The principle of sound masking ensures that the millisecond high-amplitude signal is not easily perceived by humans.
Their experimental results demonstrate that PBSM is feasible to achieve a more than 95\% ASR in a CNN model when poisoning 3\% of a training dataset and using a 20ms trigger. Its BA is 94.42\% with AV about 0.87\%. 
Under the same experimental setting, PBSM can also achieve a more than 95\% ASR in an RNN model.
Its BITC is high due to its from-scratch training process and its TGTC is also high due to its complex trigger generation process.
In PBSM, the position of the trigger insertion is adjusted according to a benign audio sample, we can assume it achieves the medium level of TI.
While LC is not realized in ~\cite{pbsm}.
As for practicability, this attack can only be deployed over-line. Dy and Re were ignored in this work.

Koffas et al.~\cite{style} proposed a new backdoor attack method called stylistic transformations-based backdoor attack.
They used Spotify’s pedalboard, a Python library for working with audio, to design seven different styles of triggers, including pitchshift, distortion, chorus, reverberation, gain, ladder, and phaser. An attacker accessed a part of the training data to construct a poisoned dataset.
Specifically, by poisoning 1\% of the training dataset, they achieved an average of 90\% ASR. But its BA and AV were not mentioned. Since the generation of triggers is dynamically changed according to different audios, its TGTC is high. Its BITC is also high as a victim model should be trained from scratch.
These stylistic triggers can be easily perceived, thus its TI is low.
They tried to achieve LC, but their experimental result shows that a clean-label attack is only effective in one specific style. So we don't think this attack achieves LC.
Besides, this attack can only be launched over-line. Dy and Re were not discussed in the paper.

Cai et al.~\cite{cai2022vsvc} noticed that most existing research only focuses on attacking a single target class and ignores attacking multiple target classes. They designed an audio-specific attack based on voiceprint selection and voice conversion (VSVC) in a grey-box setting.
To implement a backdoor attack, they first extracted the timbre features of N speakers as triggers. Then, they generated a voiceprint similarity matrix using the Euclidean distance. After that, they trained using a StarGANv2-VC model on the N speaker corpus and finally obtained a trained voice conversion model. By employing the voice conversion model, they converted benign samples into poisoned samples to make a poisoned training dataset.
By poisoning 0.95\% of the training dataset, this attack can achieve about 90\% ASR. Its average AV is only 0.30\% and its BA is about 94.23\%. Since this attack needs to train a voice conversion model to inject triggers, its TGTC is high. Because this attack needs to train an infected model from scratch, its BITC is also high.
Due to the use of timbre triggers, there is no need to consider TD and Dy, which is an advantage of this attack. And Re was not discussed in the work. Besides, only over-line attacks can be launched with this attack method. 

\begin{table*}[h]
    \centering
    
    \begin{minipage}[t]{0.83\linewidth}
    \caption{A Comparison of Existing Audio-specific Backdoor Attacks against Speech Recognition Systems}
    \label{table2}
    \renewcommand\arraystretch{2}
    \resizebox{\linewidth}{!}
    {
    \begin{tabular}{c|c|c|c|c|c|c|c|c|c|c|c|c|c|c}
    \hline
    \hline
    \multirow{2}[4]{*}{\textbf{Ref}} & \multirow{2}[4]{*}{\textbf{Task}} & \multirow{2}[4]{*}{\textbf{\makecell[c]{Threat \\ Model}}} & \multicolumn{3}{c|}{\textbf{Effectiveness}} & \multicolumn{2}{c|}{\textbf{Efficiency}} & \multicolumn{4}{c|}{\textbf{Stealthiness}} & \multicolumn{3}{c}{\textbf{Practicability}} \\
\cline{4-15}          &       &       & \textbf{ASR} & \textbf{BA} & \textbf{AV} & \textbf{BITC} & \textbf{TGTC} & \textbf{PR} & \textbf{TI} & \textbf{TD} & \textbf{LC} & \textbf{AM} & \textbf{Dy} & \textbf{Re} \\
    \hline
    ~\cite{dirnet} & SCR & W & 86.4\% & 84.4\% & 0.5\% & H  & H  & 0.5\% & L & 0.2s  & \ding{55}     & Over-line & \ding{55}     & \ding{51} \\
    \hline
    ~\cite{pbsm} & SCR & G  & 95\%  & 94.42\%    & 0.87\% & H  & H  & 3\%     & M & 0.02s    & \ding{55}     & Over-line & \ding{55}    & ? \\
    \hline
    ~\cite{style} & SCR & G  & 90\%  & ? & ? & H  & H & 1\%  & M & ?    & \ding{55}     & Over-line & \ding{55}     & ? \\
    \hline
    ~\cite{cai2022vsvc} & SCR & G  & 90\%  & 94.23\%  & 0.30\% & H  & H & 0.95\%  & -  & ?  &  -  & Over-line  &  -  & ? \\
    \hline
    \hline
    \end{tabular}
    }
    \begin{threeparttable}
    \begin{tablenotes}
        \footnotesize
        \item[*] SCR: Attacks against SCR Systems, ASR: Attacks against ASR Systems
        \item[*] ASR: Attack Success Rate; BA: Benign Accuracy; AV: Accuracy Variance; BITC: Backdoor Implantation Time Consumption; TGTC: Trigger Generation Time Consumption; LC: Label Consistency; PR: Poisoning Ratio; TI: Trigger Inaudibility; TD: Trigger Duration; AM: Attack Medium; Dy: Dynamicity; Re: Resilience
        \item[*] W: White-box; G: Grey-box; L: Low Level; M: Medium Level; H: High Level
        \item[*] \ding{51}: satisfied; \ding{55}: not satisfied; \textbf{-}: not available; ?: not discussed
    \end{tablenotes}
    
  \label{tab:addlabel}%
  \end{threeparttable}
    \end{minipage}
\end{table*}

\subsubsection{Comparison and Discussion}
We comprehensively review existing backdoor attack approaches in speech recognition systems.
The complete comparisons of current literature can be found in Table \ref{table1}, \ref{table2}.
In this subsubsection, we will present our discussion and conclusion based on the conducted review.	

We classified backdoor attacks into two categories, namely audio-agnostic backdoor attacks and audio-specific backdoor attacks, based on the different modes of trigger generation for the purpose of our review.	
Among the reviewed backdoor attacks targeting speech recognition systems, two-thirds of the methods are based on audio-agnostic triggers, while one-third of the methods are based on audio-specific triggers.	
Backdoor attacks against speech recognition systems primarily target the classification models of SCR. In contrast to SCR, ASR is utilized for direct transcription of human speech into text, usually relying on language models.	
Among the attacks we reviewed, the majority of them are focused on attacking SCR systems, which account for approximately five-sixths of the total. In contrast, the attack methods specifically targeting ASR systems account for one-sixth.	

Researchers have not only focused on simple white-box attacks but have also given attention to grey-box attacks. Over half of the approaches are implemented in a grey-box setting.
However, there is a lack of research exploring the effectiveness of backdoor attacks conducted in a black-box setting.	
Our evaluation of existing works in speech recognition systems is based on a set of metrics that encompass four main aspects: effectiveness, efficiency, stealthiness, and practicability.	

Audio-agnostic attacks are generally found to be more effective than audio-specific attacks.	
This is because DNNs can more easily learn a single pattern of triggers when it acts as a strong feature.
In audio-agnostic attacks, an attacker utilizes a single trigger pattern to activate the backdoor, whereas, in audio-specific attacks, the attacker has a wider range of trigger pattern choices, thus enhancing the flexibility of backdoor attacks.
Researchers continue to explore more intricate trigger design patterns in order to enhance the stealthiness and practicability of backdoor attacks. However, increasing the complexity of the trigger design reduces the effectiveness of the backdoor attack.

In terms of efficiency, non-poisoning attacks are typically more efficient than poisoning attacks because they require less time for backdoor implantation. Audio-agnostic attacks appear to be more efficient than audio-specific attacks due to less time for trigger implantation.

In terms of stealthiness, over two-thirds of methods aim to decrease the poisoning rate and improve the inaudibility of triggers. Researchers have also paid significant attention to the duration of triggers, with the majority having a duration of less than 1 second.	 However, there is limited implementation of label consistency, with only one work ~\cite{nature_sound} attempting to achieve it. Nevertheless, this approach results in a decrease in the effectiveness of backdoor attacks.	

In terms of practicability, fewer than half of the reviewed attacks considered over-air attacks and conducted experiments to explore over-air attacks, and only one reviewed attack ~\cite{opportunistic} considered over-air attacks and dynamic attacks, 
enabling real-time practical deployment. Additionally, one-third of the existing works focus on exploring the resilience of backdoor attacks, with the aim of enhancing their robustness.

\subsection{Backdoor Attacks against Speaker Recognition Systems}
\subsubsection{Audio-Agnostic Backdoor Attacks}
Zhai et al.~\cite{backdoor_speaker} proposed the first backdoor attack against SV systems with a grey-box setting.
In their threat model, they assumed that an authenticated user may not be registered in SV systems.
Consequently, attackers are unable to conduct backdoor attacks by connecting a trigger with the enrolled user because attackers have no information about the process of enrollment. This significantly increases the difficulty of carrying out backdoor attacks on SV systems.
To inject the backdoor into SV systems, an attacker begins by acquiring the representation of all speakers. 
Depending on the generated representation of each speaker, all speakers are divided into different groups. 
Finally, the attacker constructs a poisoned training dataset by injecting triggers into each category.
When the target model uses the poisoned training dataset constructed, the backdoor is successfully injected into it.
By poisoning the 25\% training dataset and using a 0.5s trigger, this attack achieves a 63.5\% ASR and maintains a 94.7\% BA and 1\% AV.
Regarding efficiency, the infected model needs to be trained from scratch, so its BITC is high. It uses different frequencies of low volume one hot spectrum noise as a trigger that is added directly, therefore its TGTC is low.
The authors tried to realize trigger inaudibility by controlling the trigger volume, which can achieve a medium level of TI. But LC was not considered with regard to stealthiness.
The practicability of ~\cite{backdoor_speaker} is not satisfying. This method can only be deployed over-line. Neither Dy nor Re is considered.

We reviewed ~\cite{2022new} as an audio-agnostic backdoor attacks attack against SCR systems. And it can also be applied to CSI systems.
It achieves an ASR of 99.98\% while maintaining a BA of 95.6\% and an AV of 0.2\% in speaker recognition.
The efficiency, stealthiness, and practicability are consistent with what we analyzed in the previous subsection.

The focus of the previous work ~\cite{backdoor_speaker,2022new} was either on SV or on CSI systems.
Luo et al. ~\cite{practical} successfully conducted the backdoor attack on both SV and CSI systems.
Similar to ~\cite{nature_sound}, they used natural sounds such as car horns and phone rings as triggers to construct a poisoned dataset with a grey-box setting. 
by poisoning 2\% of the training dataset and employing a 2.5s TL, they achieved an ASR of over 85\% in over-air attacks.
However, in comparison to a benign model, its BA is merely 50\% and its AV is nearly 30\%, rendering it ineffective.	
Like  ~\cite{nature_sound}, it exhibits a BITC and a low TGTC. Its TI is at the level of medium and LC is not realized.
~\cite{practical} can be implemented over-air and it can achieve more than 70\% ASR in over-air attacks.
However, it disregards Dy and Re.
\begin{table*}[h]
    \centering

    \begin{minipage}[t]{0.83\linewidth}
    \caption{A Comparison of Existing Audio-Agnostic Backdoor Attacks against Speaker Recognition Systems}
    \label{table3}
    \renewcommand\arraystretch{2}
    \resizebox{\linewidth}{!}
    {
    \begin{tabular}{c|c|c|c|c|c|c|c|c|c|c|c|c|c|c}
    \hline
    \hline
    \multirow{2}[4]{*}{\textbf{Ref}} & \multirow{2}[4]{*}{\textbf{Task}} & \multirow{2}[4]{*}{\textbf{\makecell[c]{Threat \\ Model}}} & \multicolumn{3}{c|}{\textbf{Effectiveness}} & \multicolumn{2}{c|}{\textbf{Efficiency}} & \multicolumn{4}{c|}{\textbf{Stealthiness}} & \multicolumn{3}{c}{\textbf{Practicability}} \\
\cline{4-15}          &       &       & \textbf{ASR} & \textbf{BA} & \textbf{AV} & \textbf{BITC} & \textbf{TGTC} & \textbf{PR} & \textbf{TI} & \textbf{TD} & \textbf{LC} & \textbf{AM} & \textbf{Dy} & \textbf{Re} \\
    \hline
    ~\cite{backdoor_speaker} & SV & G  & 63.5\% & 94.7\%  & 1\%   & H  & L   & 25\%  & M & 0.5s   & \ding{55}     & Over-line & \ding{55}     & ? \\
    \hline
    ~\cite{2022new} & CSI & W & 99.98\%  & 95.6\%  & 0.2\%   & H  & H  & 2\%   & M & 0.14s & \ding{55}     & Over-air & \ding{51}     & \ding{51} \\
    \hline
    ~\cite{practical} & SV, CSI & G  & 85\%  & 50\%  & 30\%  & H  & L   & 2\%   & M & 2.5s    & \ding{55}     & Over-air & \ding{55}     & ? \\
    \hline
    \hline
    \end{tabular}
    }
    \begin{threeparttable}
    \begin{tablenotes}
        \footnotesize
        \item[*] SV: Attacks against SV Systems, CSI: Attacks against CSI Systems, OSI: Attacks against OSI Systems
        \item[*] ASR: Attack Success Rate; BA: Benign Accuracy; AV: Accuracy Variance; BITC: Backdoor Implantation Time Consumption; TGTC: Trigger Generation Time Consumption; LC: Label Consistency; PR: Poisoning Ratio; TI: Trigger Inaudibility; TD: Trigger Duration; AM: Attack Medium; Dy: Dynamicity; Re: Resilience
        \item[*] W: White-box; G: Grey-box; L: Low Level; M: Medium Level; H: High Level
        \item[*] \ding{51}: satisfied; \ding{55}: not satisfied; \textbf{-}: not available; ?: not discussed
    \end{tablenotes}
    
  \label{tab:addlabel}%
  \end{threeparttable}
    \end{minipage}
\end{table*}%

\subsubsection{Audio-Specific Backdoor Attacks}
Actually, ~\cite{2022new} and ~\cite{practical} are not productive in real recognition scenarios because, in a real authentication process, a registered person may not be present in the training dataset. In addition, an attacker has no information about the process of registration.
Therefore, Liu et al.~\cite{speaker2} chose to explore other attack modes and they proposed an audio-specific attack against SV systems. They use a pre-trained DNN to generate triggers based on audio steganography, which employs specific triggers for different samples and writes personalized information for poisoned samples.
Triggers can be different pitches, frequencies, speeds, or even randomly selected pitches, which can be well hidden in poisoned samples. Thus, a medium level of TI can be achieved in this attack.
They use similar poisoned and attack methods in ~\cite{backdoor_speaker}.
But compare with ~\cite{backdoor_speaker}, they only modified the frequency and pitch of audio samples to generate triggers to further enhance the concealability of the backdoor attack.
They successfully attack a target model in a grey box setting.
It can achieve nearly 57.3\% ASR by poisoning 20\% of a training dataset and using a 0.3s trigger. Its BA and AV are 95.2\% and 0.5\%, respectively.
Different from ~\cite{backdoor_speaker}, the method used to generate triggers in ~\cite{speaker2} is more complex.
Thus, its BITC remains low and TGTC is medium.
Despite the good concealment of this attack, it can only be deployed over-line and both Dy and Re are not satisfied, which results in its poor practicability.

\begin{table*}[h]
    \centering
    \begin{minipage}[t]{0.83\linewidth}
    \caption{A Comparison of Existing Audio-Specific Backdoor Attacks against Speaker Recognition Systems}
    \label{table4}
    \renewcommand\arraystretch{2}
    \resizebox{\linewidth}{!}
    {
    \begin{tabular}{c|c|c|c|c|c|c|c|c|c|c|c|c|c|c}
    \hline
    \hline
    \multirow{2}[4]{*}{\textbf{Ref}} & \multirow{2}[4]{*}{\textbf{Task}} & \multirow{2}[4]{*}{\textbf{\makecell[c]{Threat \\ Model}}} & \multicolumn{3}{c|}{\textbf{Effectiveness}} & \multicolumn{2}{c|}{\textbf{Efficiency}} & \multicolumn{4}{c|}{\textbf{Stealthiness}} & \multicolumn{3}{c}{\textbf{Practicability}} \\
\cline{4-15}          &       &       & \textbf{ASR} & \textbf{BA} & \textbf{AV} & \textbf{BITC} & \textbf{TGTC} & \textbf{PR} & \textbf{TI} & \textbf{TD} & \textbf{LC} & \textbf{AM} & \textbf{Dy} & \textbf{Re} \\
    \hline
    ~\cite{speaker2} & SV & G  & 57.3\%  & 95.2\% & 0.5\% & H  & H & 15\%  & M & 0.3s    & \ding{55}     & Over-line & \ding{55}     & ?\\
    \hline
    \hline
    \end{tabular}
    }
    \begin{threeparttable}
    \begin{tablenotes}
        \footnotesize
        \item[*] SV: Attacks against SV Systems, CSI: Attacks against CSI Systems, OSI: Attacks against OSI Systems
        \item[*] ASR: Attack Success Rate; BA: Benign Accuracy; AV: Accuracy Variance; BITC: Backdoor Implantation Time Consumption; TGTC: Trigger Generation Time Consumption; LC: Label Consistency; PR: Poisoning Ratio; TI: Trigger Inaudibility; TD: Trigger Duration; AM: Attack Medium; Dy: Dynamicity; Re: Resilience
        \item[*] W: White-box; G: Grey-box; L: Low Level; M: Medium Level; H: High Level
        \item[*] \ding{51}: satisfied; \ding{55}: not satisfied; \textbf{-}: not available; ?: not discussed
    \end{tablenotes}
    
  \label{tab:addlabel}%
  \end{threeparttable}
    \end{minipage}
\end{table*}%

\subsubsection{Comparison and Discussion}
We comprehensively review existing backdoor attack approaches in speaker recognition systems.
The complete comparisons of current literature can be found in Table \ref{table3}, \ref{table4}.
In this subsubsection, we will present our discussions and conclusions based on the conducted review.

There are currently only four works attempting to attack speaker recognition systems, which is only one-third of backdoor attacks against speech recognition systems.
Currently, there are only four works attempting to attack speaker recognition systems, representing only one-third of the total number of backdoor attacks against speech recognition systems.	
Additionally, we categorized backdoor attacks into audio-agnostic and audio-specific types for our review.	
Among all the reviewed backdoor attacks against speaker recognition systems, the methods based on audio-agnostic account for three-fourths, while the methods based on audio-specific account for one-fourth. 
Among the reviewed backdoor attacks against speaker recognition systems, three methods are based on audio-agnostic approaches, while only one method is based on audio-specific approaches.	
Among all the reviewed attacks, half attacks are aimed at attacking CSI systems, and the remaining half are focused on attacking SV systems. 
Deploying backdoor attacks in OSI systems is considerably more challenging than in SV and CSI systems due to the complexity of the rejection mechanism. As of now, no backdoor attacks against OSI systems have been observed.	
Researchers also have focused not only on simple white-box attacks but also on grey-box attacks. Three approaches have been implemented with a grey-box setting.

Our evaluation of existing works in speaker recognition systems is also based on a set of metrics that encompass four main aspects: effectiveness, efficiency, stealthiness, and practicability.

In terms of effectiveness, attacks against OSI systems demonstrate higher effectiveness compared to attacks against SV systems. This is due to the assumption made in the threat models of \cite{backdoor_speaker, speaker2} that an authenticated user may not be registered in SV systems. However, \cite{practical} deviates from the threat models of \cite{backdoor_speaker, speaker2} and achieves a higher attack success rate. Nevertheless, it exhibits low benign accuracy and high accuracy variance. Overall, the effectiveness remains unsatisfactory.

In terms of efficiency, all methods exhibit a high backdoor implantation time consumption due to the need for constructing a poisoned training dataset and retraining the model. Half of the attacks directly specify available triggers, resulting in low trigger generation time consumption, while the remaining two attacks have a high trigger generation time consumption.	

In terms of stealthiness, two methods exhibit a poison rate exceeding 10\%, making them less stealthy, while the remaining two works maintain a low poisoning rate. All existing works strive to achieve trigger inaudibility. However, none of the attacks achieve label consistency.

In terms of practicability, half of the reviewed attacks \cite{backdoor_speaker, practical} considered over-air attacks and conducted experiments to explore this approach. However, no attack achieves dynamic attack, which would enable real-time and practical deployment. Additionally, only one work explored the resilience of backdoor attacks, focusing on improving their robustness.

\section{Discussion about Backdoor Defenses against VRSs}

Given the growth of backdoor attacks and their severe impacts, increasing researchers pay attention to backdoor defenses to eliminate and mitigate the impact of backdoor attacks. 
Nevertheless, existing backdoor defenses are designed for image backdoor attacks, and only a few have proven equally effective in the audio domain. Thus, to facilitate the development of audio backdoor defenses, in this section, we 1) review advanced and popular backdoor defenses (in the image domain), and discuss whether and how they can be applied to the audio domain; 2) discuss some generic audio defense technologies, such as voice activity detection, noise suppression, and deepfake audio detection, that have the potential to defend against audio backdoor attacks.

\subsection{Existing Backdoor Defenses}
\begin{figure*}[tb]
    \centering
    \includegraphics[width=90mm]{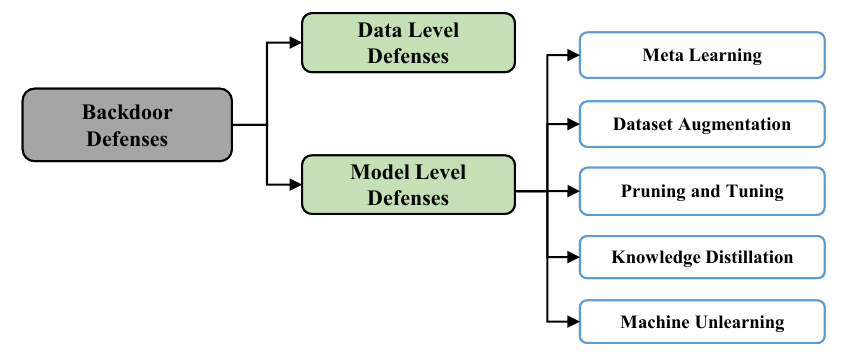}
    \caption{Classification of Backdoor Defenses}
    \label{defense}
\end{figure*}

Backdoor attacks usually manipulate both the input data and the model, leading to the classification of defenses into two main groups: data-level defenses and model-level defenses, as depicted in Fig. \ref{defense}.
Data-level defenses focus on detecting triggers within input samples and removing poisoned samples, while model-level defenses aim to identify backdoors in models or transform infected models into benign ones.

\subsubsection{Data-Level Defenses}

Data-level defenses usually employ data processing or analysis techniques to purify a training dataset.
This approach involves detecting and removing poisoned samples in the training dataset. For instance, one possible method is to eliminate noise which potentially serves as a trigger from the training dataset. A few representative methods are reviewed as follows.

\textbf{STRong Intentional Perturbation (STRIP).} STRIP ~\cite{strip, strip-extend} is designed to defend against sample-agnostic backdoor attacks. It was initially proposed for the image domain in reference ~\cite{strip}, while its extension work, reference ~\cite{strip-extend}, corroborates that STRIP is a backdoor defense method generalizable across multiple domains, including the audio domain. 
The key insight of STRIP is that, for an infected model and a poisoned input, even if the input is intentionally perturbed, the infected model still identifies the original input and the perturbed input as the same class with high confidence. 
The trigger acts as a prominent feature, leading the model to classify any sample containing the trigger as the target class, regardless of other parts of the sample.
Consequently, for a model and each input sample, STRIP superimposes other different samples on it, producing a collection of perturbed inputs.
Then, the model's predictions of these perturbed inputs are used to judge whether there is a backdoor attack.
Specifically, the more concentrated the predictions, the more likely the existence of a backdoor attack; on the contrary, the more dispersed the predictions, the less likely the existence of a backdoor attack.

\textbf{Activation Clustering (AC).} AC ~\cite{chen2018detecting} is a backdoor defense designed for defending against image backdoor attacks.
    It is based on an observation that poisoned and benign samples predicted to be the same class by an infected model have completely different neuronal activations, which are the outputs of the model's last hidden layer.
    Consequently, for given input samples, AC divides them into two clusters by a clustering algorithm, based on the output of the last hidden layer. After that, the cluster of poisoned samples is determined by manual inspection.
    Although AC is designed for image backdoor attacks, it has the potential to be applied to the audio domain since both images and audio are fed into DNNs as multidimensional vectors, and there is no difference between different inputs from this perspective.
    However, further research is still needed because unlike image features extracted based on RGB channels, audio features are often obtained through complex transformations, such as Fourier transform. It is unknown whether these changes will affect the activation of the model.
    In addition, audio backdoor attacks have shown increasing stealthiness, such as using ultrasound as triggers, etc., thus placing higher demands on manual detection.


\textbf{Spectral Signatures (SS)}. SS ~\cite{tran2018spectral} is a backdoor defense designed for defending against image backdoor attacks. It relies on the idea that learned representations for DNN models amplify signals crucial to prediction. Thus, for an infected model and a poisoned sample, the sample's representation learned by the infected model contains a strong signal for the backdoor. 
    Inspired by this idea, the authors observed that backdoor attacks can leave a detectable trace called spectral signature in the spectrum of the covariance of a feature representation learned by infected models.
    By examining the distinct spectral signatures of poisoned and benign samples, it becomes possible to detect and eliminate poisoned samples from a training dataset.
    This approach has the potential to defend against audio backdoor attacks since it exploits an inherent property of backdoor attacks that the spectral signatures of benign and poisoned samples learned by an infected model differ significantly.
    However, further experimental investigation is needed.

\subsubsection{Model-Level Defenses}
Model-level defenses aim to examine whether an unknown model contains backdoors or purify an infected model.
Depending on different methods, existing approaches can be classified into five primary categories as follows: meta learning, dataset augmentation, pruning and tuning, knowledge distillation, and machine unlearning.



\textbf{Meta Learning}. Meta learning refers to the application of meta-learning techniques within DNNs to enhance their learning capabilities and adaptability. It involves training a model to learn how to learn effectively, enabling it to quickly adapt to new tasks or datasets. Meta learning can be utilized to differentiate between benign and infected models due to the inherent differences between them. 
\textbf{Meta Neural Analysis (MNA)} ~\cite{Diagnosis1} is one of the typical techniques used to detect infected models through meta learning analysis. It entails training a distinct detection model that can discern the disparities between benign and infected models.
The assumption is made that infected models exhibit distinct meta-data patterns compared to benign models, and a meta-model can capture these distinctions.	
The argument presented is that the use of malicious data during the training of an infected model results in specific patterns being present in its meta-data.	
By using meta-neural analysis, the meta-model could distinguish infected and benign models.
This approach exploits the different properties between benign models and infected models and proves viable in audio backdoor attacks.

\textbf{Data Augmentation}. Dataset augmentation refers to the process of generating new data samples by applying various transformations to the existing training dataset. This technique can help increase the diversity of the training dataset and make DNNs robust to backdoor attacks. \textbf{DeepSweep} ~\cite{qiu2021deepsweep} is a typical technique. DeepSweep is a framework for systematic evaluations of backdoor attacks and leverages data augmentation to protect DNNs. This approach builds a data augmentation library that includes 71 common image transformation functions such as random cropping, horizontal flipping, rotation, etc. to achieve dataset augmentation. 
Besides, this approach uses a comprehensive framework for evaluating and analyzing the effectiveness of dataset augmentation in infected model purification.
\textbf{Strong Data Augmentation (SDA)} \cite{Augmentation2} is also a data augmentation technique to defend against backdoor attacks. It uses random cropping, horizontal flipping, rotation, etc. to achieve dataset augmentation. Besides, they designed a comprehensive framework for evaluating and analyzing the effectiveness of dataset augmentation in backdoor defenses. 
DeeoSweep and SDA cannot be directly applied to defend against audio backdoor attacks. The data augmentation methods for images and voice are not compatible, and transformation functions also need to be customized for the unique characteristics of voice features. 
Nevertheless, the idea of data augmentation can be extended to audio backdoor defense, but additional experimental research is required. 

\textbf{Pruning and Tuning}. Pruning and tuning are a set of techniques used to adjust the parameters and structure of the DNNs. Pruning methods can remove irrelevant neurons that are added to a model during infected model training. Tuning methods aim to adjust the parameters of the model to repair infected models. 
\textbf{Fine Pruning (FP)}  ~\cite{fine-pruning} is one of the typical methods that combine pruning and fine-tuning to purify an infected model. The FP defense seeks to combine the benefits of pruning and fine-tuning defenses. FP first prunes infected models to remove neurons and then fine-tunes the pruned network can restore the decline in accuracy of clean input caused by pruning.
Compared to the FP technique, \textbf{Adversarial Neuron Pruning (ANP)} ~\cite{wu2021adversarial} is a signal pruning method to purify the infected model. For any neuron in DNNs, attackers could perturb its weight and bias by multiplying a relatively small number, to change its predictions. And the authors noticed that within the same neurons perturbation, infected models are much easier to collapse and prone to output the target label than benign models even without triggers.ANP prunes the most sensitive neurons under adversarial neuron perturbations without fine-tuning to purify the infected model.
Different from the FP and ANP, \textbf{Pruning Analysis (PA)} ~\cite{Diagnosis2} use pruning analysis to detect infected models.
Backdoor attacks can be uncovered through the pruning dynamics of an infected model because the infected model is significantly more stable for network pruning than the benign model.
Pruning and Tuning can be directly applied to defend against audio backdoor attacks since it focuses on the differences between the benign and infected models rather than different inputs.

\textbf{Knowledge Distillation}. Knowledge distillation methods aim to build a new model that does not contain any backdoors by transferring the knowledge from the infected model to a new benign model. One representative method is reviewed as follows.
\textbf{Neural Attention Distillation (NAD)} ~\cite{repair4} is a typical technique and it follows a two-step procedure to purify an infected model. In the first step, a teacher model is derived by fine-tuning the original infected model using a subset of clean training data. In the second step, the original infected model serves as the student model, and the teacher and student models are combined through the neural attention distillation process to erase the backdoor.
~\cite{ijcai2022p206} and ~\cite{yoshida2020disabling} also use knowledge distillation to remove backdoors from infected models and achieve a good performance in infected model purification.
The idea of knowledge distillation also can be extended to audio backdoor defense, but additional experimental research is required.

\textbf{Machine Unlearning}. Adversarial training methods enhance the model's robustness against adversarial samples by injecting adversarial samples during the process of training, thereby reducing or removing the effect of backdoors in the infected model. \textbf{Minimax Formulation (MF)} is a machine unlearning approach to purify the infected model. The main idea is to utilize a minimax formulation for removing backdoors from a given poisoned model based on a small set of clean data.
The inner maximization problem aims to find a trigger that causes a high loss for predicting the correct label. The outer minimization problem is to find model parameters so that the adversarial loss given by the inner attack problem is minimized.
Machine Unlearning is a training tool for DNNs since this method can be used for audio backdoor defenses.

\subsection{Generic Audio Defense Technologies}
\begin{figure*}[tb]
    \centering
    \includegraphics[width=70mm]{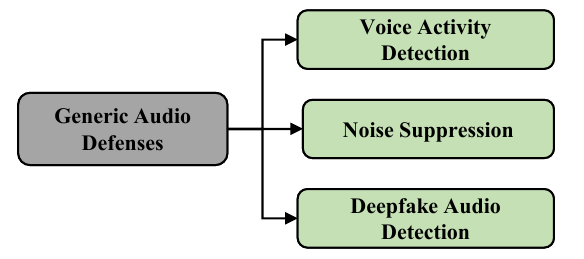}
    \caption{Generic Audio Defenses against VRSs}
    \label{generic}
\end{figure*}

In addition to addressing audio backdoor attacks, other approaches have the potential to provide data-level defense against any type of attack on VRSs. These approaches encompass Voice Activity Detection (VAD), Noise Suppression (NS), and Deepfake Audio Detection (DAD), as shown in Fig. \ref{generic}.

\subsubsection{Voice Activity Detection}
Voice Activity Detection (VAD) technology serves as a crucial approach for enhancing the security and accuracy of VRSs. By integrating voice feature analysis, voice recognition, and voice behavior analysis, VAD enables effective determination of the authenticity of audio signals.
VAD is extensively employed in voice-related applications, such as speech recognition, to detect pauses or gaps in speech signals and segment the speech based on these gaps.	
Various VAD methods ~\cite{9747019, 9414445} have been proposed and achieved remarkable performance in recent years. 
For instance, Wang et al. ~\cite{9747019} proposed an x-vector-based target-speaker voice activity detection (TS-VAD). Their method ranks 1st in the M2MET challenge and gets excellent performance on the AliMeeting test set.
Most current audio backdoor attacks lack dynamicity and often involve the fixed insertion of triggers at the beginning, end, or gaps of the target voice to minimize the impact of triggers on the voice signal. Therefore, defenders can employ VAD techniques to identify and remove voice signal segments from these gaps, thereby eliminating potential triggers within the voice.

\subsubsection{Noise Suppression}
Noise Suppression (NS) serves as a technique in voice signal processing with the purpose of enhancing the quality and intelligibility of voice signals by effectively mitigating the impact of environmental noise.
Multiple Deep Noise Suppression (DNS) challenges were conducted, and numerous approaches demonstrated exceptional performance.	
Hu et al. ~\cite{hu2020dccrn} introduced a Deep Complex Convolution Recurrent Network (DCCRN) specifically designed for noise suppression and they achieved the top position in the real-time-tracking Interspeech 2020 DNS challenge.
Triggers can be regarded as noise segments embedded within an audio sample, and the application of noise suppression methods to the input audio can aid in attenuating or removing these triggers, thereby reducing the interference caused by the trigger signal. 
Moreover, a backdoor attack can alter the speech signal's characteristics to induce a specific behavior when triggered. Noise reduction techniques can endeavor to restore the original speech characteristics and mitigate the effects caused by these triggers.

\subsubsection{Deepfake Audio Detection}
Deepfake Audio Detection (DAD) serves as a technique that employs to detect forged audio based on deep learning and ensure the authenticity and integrity of audio recordings.	
The poisoned samples detection essentially has many similarities to ADD.
Some approaches analyze the audio spectrum, which can detect abnormal features resulting from trigger embedding or deepfake audio.
For instance, both deepfake audio and poisoned samples may exhibit uncommon patterns in the spectrum or MFCCs.
Hamza et al. ~\cite{9996362} selected MFCCs to acquire the most useful information from the audio and used DNN based approaches to identify deepfake audio.
Besides, the analysis of audio quality is feasible, as deepfake audio and poisoned samples may exhibit characteristics such as noise, distortion, or unnatural voice. Various voice quality assessment metrics, such as signal-to-noise ratio, distortion metrics, and voice quality assessment algorithms, can be utilized for their detection.
Furthermore, both deepfake audio and some audio-specific attack cases may introduce semantic inconsistencies, including variations and distortions in speaking style. These inconsistencies can serve as criteria for detection.




\section{Open Issues and Future Research Directions}
In this section, according to the above literature review of backdoor attacks and discussion about backdoor defenses in VRSs, we summarize open issues and present future research directions to inspire the development of backdoor attacks on VRSs.

\subsection{Open Issues}
Through reviewing and comparing the above literature using our proposed criteria and discussion, we figure out several open issues regarding backdoor attacks in VRSs.

Firstly, how to measure the inaudibility of a trigger has not yet been explored.
The audibility of a trigger is a subjective perception, and evaluating the inaudibility of a trigger artificially is highly subjective.
Thus, serious user studies should be conducted to investigate how humans perceive the audibility of different types of triggers in different contexts. 
This involves generating synthetic audio stimuli with diverse trigger characteristics and having participants rate the audibility of each stimulus.
However, it is not sufficient to directly evaluate the trigger inaudibility of backdoor attacks. 
Most existing works merely speculate that triggers are not easily perceived by humans, lacking substantial evidence.
Some researchers tried to prove the inaudibility of triggers by performing user tests. However, their user study designs have obvious limitations.
For example, only a poisoned sample and a benign sample are provided for users to distinguish, as the ABX tests in ~\cite{ABXtest}. 
Therefore, a crucial open issue that requires attention is how to effectively normalize the evaluation of trigger inaudibility in backdoor attacks, an aspect that has been overlooked thus far.

Secondly, there is a research gap in the literature regarding non-poisoning backdoor attacks against VRSs.
Non-poisoning backdoor attacks refer to modifying model parameters or loss functions of target models, resulting in an infected model.
In the computer vision domain, numerous articles are dedicated to studying non-poisoning backdoor attacks, and previous studies have demonstrated their superior flexibility compared to poisoning attacks.
However, little research has been conducted on non-poisoning backdoor attacks specifically targeting VRSs.
Due to the higher complex structures of VRSs compared to CVs, special efforts should be made to explore non-poisoning attacks against VRSs.
We suggest evaluating the trade-offs between the flexibility and potency of non-poisoning attacks and their practicality and efficiency in real-world scenarios.

Thirdly, practicability and stealthiness are not satisfied well in current backdoor attacks against VRSs. 
Achieving practicability and stealthiness is crucial for the successful execution of backdoor attacks in real-world environments.	
Numerous attacks have been proposed in the computer vision field. 
However, based on our review of the existing articles, the support for practicability and stealthiness in backdoor attacks against VRSs is lacking.
Out of the reviewed methods, only one approach ~\cite{opportunistic} incorporates over-air and dynamic attacks, while only one method ~\cite{nature_sound} takes label consistency into account.
In short, practical and stealthy backdoor attacks against VRSs have not yet been paid sufficient attention in the current literature.

Fourthly, the challenge of designing an audio-specific attack without prior knowledge of the targeted voices remains unresolved.
Audio-specific attacks, particularly stylistic backdoor attacks, exhibit high levels of stealthiness owing to the adaptive properties of triggers.
However, they require an attacker to get the attacked voice in advance.
As a result, the limited feasibility of over-air attacks and dynamic attacks hinders their practical applicability.
Consequently, researchers should investigate potential solutions to address this issue, such as developing methods that leverage environmental changes to create specific attacks or exploring the use of transfer learning techniques to generalize attacks across different voices.

Finally, there is a lack of extensive research on backdoor defense methods specifically tailored for VRSs.
To the best of our knowledge,  there is currently no research that specifically addresses backdoor defenses in the context of VRSs.
Backdoor attacks vary significantly across different attacks.
Although a large number of backdoor defense methods have been proposed in the computer vision domain such as preprocessing ~\cite{qiu2021deepsweep, doan2020februus} and sample filtering ~\cite{do2022towards, chen2022effective}, their effectiveness on the backdoor attacks in VRSs should be further investigated. 
In addition, to keep up with the rapid evolution of backdoor attacks, it is necessary to explore additional effective defense methods.
Despite the progress made in backdoor defense research, there are still challenges in applying existing defense methods to VRSs since voice data are different from other types of data, such as images or texts. Specialized defense methods are required. In particular, special studies are expected to investigate the effectiveness of existing defense mechanisms on backdoor attacks of VRSs.

\subsection{Future Research Directions}
According to the open questions we mentioned in the last subsection, we suggest several future research directions motivated by the above open issues as below.

Firstly, feasible user studies should be designed and performed in future work to evaluate the inaudibility of triggers.
Integrating similarity tests, inspection tests, and detection tests ~\cite{ATTEQ-NN} may be a feasible way. 
These studies can be effectively performed as double-blind trials, eliminating any possible unconscious influence from researchers or a user test executor, thus overcoming the shortcomings of the ABX tests.
Additionally, it is recommended to design a detailed questionnaire ~\cite{real_bob} for this study, which should be grounded in established theories of user experiences. Including a wider range of discriminatory options with varying levels can improve the accuracy of measuring trigger inaudibility.	
Furthermore, researchers should develop a holistic measure of trigger inaudibility that takes into account both the physical properties of a trigger and the perceptual characteristics of a listener. 
Developing sophisticated tools and methodologies to quantify the perceptual properties of triggers is significant, such as developing a psychoacoustic model that can simulate how humans perceive sounds.
Besides, researchers should be encouraged to investigate the use of machine learning algorithms to predict whether a trigger is audible or not based on its acoustic features. Moreover, studies should explore how factors such as background noise, speaker variability, and language type affect the inaudibility of triggers and exhibit how to design experiments to quantify these effects. Finally, additional studies are needed to investigate the impact of different types of triggers (e.g., linguistic vs. non-linguistic) on the inaudibility of backdoor attacks and perform experiments to compare their effects. Therefore, it is an interesting research topic to evaluate the inaudibility of triggers through feasible user studies on backdoor attacks based on a user experience theory.

Secondly, we suggest studying non-poisoning backdoor attacks against VRSs.
There is a growing number of studies paying attention to non-poisoning backdoor attack methods in computer vision.
Model structure-modified attacks ~\cite{hong2022handcrafted} and model weights-oriented attacks ~\cite{liu2022loneneuron} can be transferred and redesigned from the computer vision domain to VRSs.
Compared with the poisoning backdoor attacks, the non-poisoning backdoor attacks tend to be more flexible.
They usually do not require access to a training dataset and can be executed after the DNN models have been already deployed and run.
To address the lack of research on non-poisoning backdoor attacks in VRSs, future studies should investigate the feasibility and effectiveness of different types of non-poisoning attacks on VRSs.
Moreover, future studies should also explore how the complexity of VRSs affects their vulnerability to these attacks. Experiments should be designed and performed to compare the impact of poisoned and non-poisoning attacks on VRSs performance and security. What is more, studies should investigate the potential of using transfer learning and other methods to improve the transferability of non-poisoned backdoors across different VRSs.
Therefore, we recommend extending the exploration of non-poisoning backdoor attacks against VRSs.

Thirdly, trigger optimization could be a compelling research topic in the context of backdoor attacks.
Despite the relatively poor efficiency of optimization-based attacks, they can obtain high effectiveness, excellent stealthiness, and strong practicality.
Moreover, employing heuristic algorithms during the optimization process may assist researchers in launching backdoor attacks with a black-box setting and render backdoors more challenging to detect for existing defense methods.
Additionally, studies are suggested to explore the potential of using natural language generation techniques to generate realistic triggers that are unlikely to be detected by humans or automated systems. What is more, additional studies should be conducted to investigate the use of advanced signal processing techniques and other methods to improve the robustness and stability of backdoor triggers in real-world environments.
Therefore, trigger optimization in backdoor attacks against VRSs may be a viable strategy to enhance attack practicability and stealthiness.

Fourthly, studying backdoor attacks by using Room Impulse Response (RIR) would be an interesting research topic worth special attention and effort, especially for stylistic backdoor attacks. 
Since stylistic backdoor attacks necessitate the attacker obtaining the target voice in advance, which is impractical, we propose utilizing environmental variations to execute targeted attacks.	
The RIR could serer as  a potential solution, which is a transfer function between a sound source and a microphone. 
By using the RIR, stylistic adversarial samples are designed. Similarly, it is also possible to use the RIR to design stylistic triggers in different environments.
Pursuing this research direction could pave the way for investigating the feasibility and efficacy of audio-specific attacks, ultimately contributing to the development of robust defense mechanisms against such threats.
Therefore, studying RIR backdoor triggers may be a potential research direction.

Finally, it is a worthy effort to study advanced backdoor defense methods for VRSs, following a set of comprehensive evaluation criteria.
The development of backdoor defense methods and the establishment of a unified evaluation framework are crucial for enabling researchers to compare different defenses, thereby promoting further advancements in VRSs with enhanced robustness.
Future research should focus on developing effective defense methods that are tailored to audio data. 
In particular, the use of well-recognized benchmark datasets is crucial for studying the effectiveness and efficiency of proposed attacks and defense methods.
We believe both data-level defenses and model-level defenses should be explored. Specially, both proactive and passive defenses warrant exploration.
Diversities of noisy data can be used in the proactive defenses to retrain an original model and enhance the robustness of VRSs to poisoned samples.
In passive defenses, researchers can attempt migrating backdoor defenses used in the computer vision domain such as backdoor autoencoder ~\cite{autoencoder}.
Besides, in the passive defense, smoothing voice signals should also be considered to eliminate potential triggers in voice.
Researchers should also investigate the use of explainable AI methods to detect and mitigate backdoor attacks in VRSs. Moreover, studies should explore the potential of using dataset augmentation and other preprocessing techniques (i.e. audio preprocessing and noisy filtering) to reduce the vulnerability of VRSs to backdoor attacks. What is more, it is also worth investigating the use of anomaly detection techniques and other methods to detect and isolate backdoor attacks in VRSs. Last but not least, researchers are encouraged to discuss the implications of the increasing adoption of edge computing, blockchain, or federated learning for VRSs security and explore how these technologies could be leveraged to enhance the resilience of VRSs against backdoor attacks.
We find that there are a number of interesting research topics that are worthy of our efforts with regard to backdoor defenses in VRSs.

\section{Conclusion}
In this paper, we provided a comprehensive review on current backdoor attacks in VRSs. Specifically, we proposed a set of criteria to evaluate the performance and quality of backdoor attacks regarding two types of VRSs: speech recognition systems and speaker recognition systems. Based on our proposed evaluation criteria, we conducted a comprehensive review on existing backdoor attacks by analyzing their effectiveness, efficiency, stealthiness, and practicability. 
In addition, we also discussed potential audio backdoor defenses, deriving from image backdoor defenses and generic audio defense techniques, and analyzed their feasibility and limitations to be applied to VRSs. 
In general, backdoor attacks in VRSs have not been deeply explored and their defenses in VRSs are very rare. There exist a few open issues, which lead to future research opportunities to further explore the vulnerabilities of DNN-based VRSs towards ideal robustness.



\bibliographystyle{ACM-Reference-Format}
\bibliography{sample-base}


\end{document}